\documentclass[twocolumn]{aastex61}
\usepackage{amsmath}
\usepackage{graphicx}
\usepackage{epsfig}
\usepackage{natbib}
\usepackage{wasysym}
\usepackage{lipsum}
\usepackage{mathrsfs}
\usepackage{float}
\usepackage{rotating}
\usepackage{graphicx}
\usepackage{epstopdf}
\usepackage{gensymb}
\usepackage{amsmath} 

\newcommand{\kepler}{{\it{Kepler}}}
\usepackage{color}

\newcommand{\gyr}{{\rm{Gyr}}}
\newcommand{\mearth}{{M_{\oplus}}}
\newcommand{\rearth}{{R_{\oplus}}}

\newcommand{\mpl}{{M_{\rm pl}}}
\newcommand{\menv}{{M_{\rm{env}}}}
\newcommand{\fpl}{{F_{\rm{pl}}}}

\newcommand{\rcr}{{R_{\rm{cr}}}}
\newcommand{\yr}{{\rm{yr}}}
\newcommand{\myr}{{\rm{Myr}}}
\newcommand{\deltazeroten}{{\delta_{0.1-0}}}
\newcommand{\deltazerofive}{{\delta_{0.05-0}}}
\newcommand{\deltazerox}{{\delta_{x-0}}}
\newcommand{\au}{{\rm{AU}}}
\newcommand{\Rpl}{R_{\rm{pl}}}

\usepackage{hyperref}
\usepackage{longtable}

\usepackage{filecontents}

\begin{document}

\bibliographystyle{apj}

\shorttitle{Planetesimal Accretion by Sub-Neptunes}
\shortauthors{Chatterjee, S. \& Chen, H.}

\title{Effects of Planetesimal Accretion on the Thermal and \\Structural Evolution of Sub-Neptunes}

\author[0000-0002-3680-2684]{Sourav Chatterjee}

\affil{ Department of Physics \& Astronomy, Northwestern University, Evanston, IL 60202, USA}
\affil{ Center for Interdisciplinary Exploration \& Research in Astrophysics (CIERA), Evanston, IL 60202, USA}

\author[0000-0003-1995-1351]{Howard Chen}

\affil{ Center for Interdisciplinary Exploration \& Research in Astrophysics (CIERA), Evanston, IL 60202, USA}
\affil{Department of Earth \& Planetary Science, Northwestern University, Evanston, IL 60202, USA}

\email{sourav.chatterjee@northwestern.edu, howardchen2021@u.northwestern.edu}

\begin{abstract}
A remarkable discovery of NASA's {\it Kepler} mission is the wide diversity in the average densities of planets 
of similar mass. After gas disk dissipation, fully formed planets could 
interact with nearby planetesimals from a remnant planetesimal disk. 
These interactions would often lead to planetesimal accretion due to the 
relatively high ratio between the planet size and the hill radius for typical planets. 
We present calculations using the open-source stellar evolution toolkit {\sc mesa} 
(Modules for Experiments in Stellar Astrophysics) modified to include the deposition 
of planetesimals into the H/He envelopes of sub-Neptunes (${\sim} 1-20\,\mearth$). 
We show that planetesimal accretion can alter
the mass-radius isochrones for these planets. 
The same initial planet as a result of the same total accreted planetesimal mass 
can have up to $\approx5\%$ difference in mean densities several ${\sim}\gyr$ after 
the last accretion due to inherent stochasticity of the accretion process.
During the phase of rapid accretion these differences are more dramatic. 
The additional energy deposition from the accreted planetesimals increase
the ratio between the planet's radius to that of the core during rapid accretion, which in turn leads to 
enhanced loss of atmospheric mass. As a result, the same initial planet can end up with very different 
envelope mass fractions. These differences manifest as differences in mean densities 
long after accretion stops. 
These effects are particularly important for planets initially less massive 
than $\sim10\,\mearth$ and with envelope mass fraction less than 
$\sim10\%$, thought to be the most common type of planets discovered by {\it Kepler}.  
\end{abstract}
\keywords{hydrodynamics--scattering--methods: numerical--planets and satellites: atmospheres--planets and satellites: physical evolution--planet-disk interactions}

\section{Introduction} 
\label{sec:intro} 

Exoplanetary observational campaigns discovered a profusion of planetary systems with a wide range of dynamical and structural properties \citep{RoweEt2014ApJ,MullallyEt2015ApJS}. Large-scale transit surveys such as {\it Kepler} transformed our understanding of the makeup of a ``normal" planet. Absent in 
the Solar System, sub-Neptune-size planets ($5~M_\oplus \la M_p \la 20~M_\oplus$; $0.3$ gm/cm$^3$ $\la \rho_p \la 1.5$ gm/cm$^3$) with short-period orbits dominate the current population of known planets, providing compelling reasons to understand how they form and evolve to their observed orbital and structural architectures \citep{Wolfgang+Lopez2015ApJ,MullallyEt2015ApJS}. 

Measured masses of the transiting planets exhibit yet another surprising trend; for any given size of the planets, 
especially in the sub-Neptune regime, the average densities vary by orders of magnitude \citep[e.g.,][]{MarcyEt2014ApJS,Weiss&Marcy2014ApJL}. Indeed, 
several studies suggested that empirical mass estimates for 
a planet of a given size should be determined 
probabilistically from distributions with large ranges rather 
than deterministically using power-laws 
\citep[e.g.,][]{Chatterjee+Tan2015ApJL,Wolfgang+Rogers+Ford2016ApJ}. It is now well understood that in this regime the mass 
of the planet is dominated by a dense core, whereas the size of the planet is determined by a low-density and low-mass-fraction gaseous envelope \citep[e.g.,][]{Weiss&Marcy2014ApJL,Wolfgang+Lopez2015ApJ,Rogers2015ApJ}. As a result, the average densities for any given planet mass (or size) are dependent on the structure of the gaseous envelopes, which in turn can depend on internal properties including the core-to-envelope 
mass fraction \citep[e.g.,][]{Rogers&Seager2010bApJ}, envelope composition \citep[e.g.,][]{Rogers&Seager2010bApJ}, 
and even the entropy profile in the envelope 
\citep[e.g.,][]{HoweEt2014ApJ}. In addition, several physical processes, external to the planet, can bring dramatic changes to the structure of the envelope. Most recently, it was shown by both models and observations that the sizes of short-period, 
sub-Neptune-mass planets may have been sculpted by irradiation from the host star via envelope evaporation  
\citep[e.g.,][]{Owen&Wu2013ApJ,Lopez&Fortney2013ApJ,JinEt2014ApJ,Chen+Rogers2016ApJ,FultonEt2017,Owen+Wu2017arXiv,lehmer2017hydrodynamic}. 

To the first order, the insolation flux and the planet's core 
mass set the planet's thermal evolution and controls mass loss 
from its envelope \citep{Lopez&Fortney2013ApJ}. However, 
other processes, dependent on the dynamical history of the 
planets may also affect their evolution and final observable 
structural properties. Multiplanet systems observed 
by \kepler\ are shown to be filled to capacity and the 
distribution of separations (in terms of their Hill radii) between adjacent planet pairs 
suggests that the currently 
observed planetary orbits may have been sculpted via past dynamical 
instabilities \citep[e.g.,][]{Fang+Margot2013ApJ,Pu+Wu2015ApJ}. Planet-planet instabilities in this regime typically 
result in physical collisions \citep[e.g.,][]{JeongAhn+Malhotra2017AJ} rather than strong scattering typical of giant planets \citep[e.g.,][]{Chatterjee+etal+2008ApJ}. Physical collisions between sub-Neptunes of course can dramatically 
alter the observable average properties, for example, by stripping of envelopes \citep[e.g.,][]{Hwang+Chatterjee+etal2017ApJ}.   

Based on the core-accretion paradigm of planet formation,
planetesimals are thought to accompany planet formation 
(e.g., \citealt{GoldreichEt2004}). We do observe relics of these in the form of the Kuiper belt objects (KBO) and the Asteroids in the Solar system \citep[e.g.,][]{Morbidelli+etal+2009Icarus}. 
It is expected that before gas-disk dispersal, fully formed
planets may be embedded in residual planetesimal disks 
stabilized by the dissipations from the gas disk. 
Absent the dissipative effects from the gas disk 
after gas dispersal (or when the gas densities are 
sufficiently low), the planets may dynamically interact with
nearby planetesimals. Indeed, planetesimal-driven 
migration, in presence or absence of a gas disk, is well studied in the past, especially in the context of the Solar system and has long been identified as an important ingredient to understand the formation and dynamical evolution of planetary systems \citep[e.g.,][]{Fernandez+Ip+1984Icarus,Hahn+Malhotra+1999AJ,Krish+etal+2009Icarus,Bromley+Kenyon+2011ApJ,Ormel+etal+2012ApJ,Minton+Levison+2014Icarus}. Most recently, it was also argued that the orbital architectures of the \kepler\ planets, more specifically, the distribution of period ratios for near-resonant planet pairs indicate that most planets may have gone through a phase of planetesimal accretion after gas-disk dispersal \citep[][]{Moore+etal+2013MNRAS,Chatterjee+Ford2015ApJ}. While the above studies focused on the orbital evolution of planets as a result of planet-planetesimal interactions, \citet{Chatterjee+Ford2015ApJ} suggested that these planets may accrete planetesimals of total mass $\sim$ a few to $10\%$ of its own mass. Thus, such high level of accretion after planet formation and gas-disk dispersal may also dramatically affect the typically low-mass envelopes on these planets and hence their observable properties. 

Earlier works have investigated the effects of heavy elements deposition into the interiors of gas giants. They argued that such deposition can lead to the suppression of the inwards growth of the convection zone, which may signiﬁcantly delay cooling \citep{Leconte+Chabrier2012,Lozovsky2017jupiter}. 
It is expected that these effects would be much greater for the lower-mass, sub-Neptune-sized planets ($M_p \la 15~M_\oplus$), typical of the \kepler\ planets. 
\citet{Inamdar+Schlichting2015} studied the effects of 
giant impacts during planet formation and argued that 
impacts can be a key cause for atmospheric escape as well 
as the delay in gas accretion until gas-disk dispersal. 
More recently, \citet{MordasiniEt2017arXiv} found that heating contributed by planetesimal accretion could produce an upturn in the planetary luminosity-mass relationships soon after formation.

In this paper we report a systematic study of the effects 
of planetesimal accretion {\it after} planet formation and 
{\it after} gas-disk dispersal. Specifically, we focus on 
whether late stage accretion of planetesimals by 
sub-Neptune-size planets can influence the planets'
thermal/mass-loss histories. 

The paper is presented as follows. In 
Section~\ref{sec:meth}, we describe the numerical setup 
for the evolution of $\sim$Neptune-mass planets. We 
present our key results showing the effects of planetesimal accretion on the structural evolution of 
planetary envelopes in 
Section~\ref{sec:results}. We discuss the implications of our results and conclude in Section~\ref{sec:discussion}.  

\section{Numerical Setup}
\label{sec:meth} 
 
We use the state-of-the-art software {\sc mesa} \citep[][version 8845]{PaxtonEt2011ApJS, PaxtonEt2013ApJS, PaxtonEt2015ApJS} to model structural evolution of 
planets. Similar to past planet evolution studies (e.g.\citealt{ValenciaEt2007ApJ,LopezEt2012ApJ,BerardoEt2017ApJ}), we consider spherically symmetric planets consisting of a heavy-element interior (comprised of rocky material) surrounded by a hydrogen-helium (H/He) dominated envelope. 

We adopt the H/He equation of state (EOS) from \citet{SaumonEt1995ApJS} for the planetary atmosphere. Unless otherwise stated, we use star-planet metallicity $Z = 0.03$ and helium mass fraction $Y = 0.25$. 
The metallicity and helium compositions of most exoplanets are unobserved, hence the assumption of near-solar values is a reasonable starting point.

We use the standard low temperature Rosseland tables \citep{FreedmanEt2008,FreedmanEt2014} for visible and infrared opacities. 
The {\sc mesa} EOS and opacity tables are further described in \citet{PaxtonEt2011ApJS, PaxtonEt2013ApJS}. Note that Freedman et al. opacities do not include dust grains. 
Hence in this first study we did not include composition changes in the planetary envelope due to planetesimal accretion. Modeling the planetesimal-envelope interaction (e.g., through fragmentation of planetesimals and likely complex planetesimal chemistry) in detail is beyond the scope of this work and will be explored in a separate study. 
Nevertheless, see Section~\ref{sec:discussion} for a discussion about how different adopted metallicities and opacity may affect a planet's structural evolution.
As per tradition, we will make our extensions available for the community at the {\sc mesa} marketplace (\url{http://cococubed.asu.edu/mesa_market/add-ons.html}).

\subsection{The Planet Model}

We closely follow the setup of \citet{Chen+Rogers2016ApJ} 
who introduced a 
self-consistent initial starting routine, atmospheric boundary conditions that are more suitable for sub-Neptunes, a refined thermal-physical model for the planet core, and a coupled thermal-evaporative evolution. 
We assume spherical symmetry 
for the planetary internal structures similar to a wide range of past 
planet evolution models (e.g. \citealt{ValenciaEt2010A&A, Owen&Wu2013ApJ,Howe+Burrows2015ApJ}).

The treatment of the atmosphere is based on the two-stream approximation of \citet{Guillot&Havel2011A&A}, where the incident upward and downward fluxes are treated separately with the assumption of a grey atmosphere independent of incident radiation wavelength. The planet interior models are derived from \citet{RogersEt2011ApJ} with mass-dependent core densities and time-dependent core luminosity. Lastly, the thermal evolution model is coupled with an analytic mass-loss prescription previously considered in detail by numerous studies \citep[e.g.,][]{WatsonEt1981,Kasting+Pollack1983Icarus,BaraffeEt2003A&A,LammerEt2003ApJ,ErkaevEt2007A&A,ValenciaEt2010A&A,LopezEt2012ApJ,salz2016energy},
\begin{equation} 
\frac{{\rm d} M_p}{{\rm d} t} = - \frac{\epsilon_{\rm EUV} \pi F_{\rm EUV} R_{\rm p} R_{\rm EUV}^2}{G M_p K_{\rm tidal}}, \label{elim}
\end{equation}

where $\epsilon_{\rm EUV}$ is the mass loss efficiency parameter and we have adopted a value of $0.15$ in this study \citep{JacksonEt2012MNRAS,LopezEt2012ApJ}. Calculation of the precise value requires knowledge of the detailed atmospheric composition and is beyond the scope of this study. $F_{\rm EUV}$ is the extreme ultraviolet energy flux from the host star impinging on the planet atmosphere. $R_{p}$ and $M_p$ are planet radius at optical depth $\tau_{\rm visible} = 1$ (in the visible) and the total mass of the planet, respectively. $G$ is the gravitational constant. $R_{\rm EUV}$ is the distance from the center of the planet to the point where the atmosphere is optically thick to EUV photons. 
$K_{\rm tidal}\equiv 1-(3R_{\rm EUV})/(2R_{H})+ 1/[2(R_{H}/R_{\rm EUV})^3]$ corrects for tidal forces, which modify the geometry of the potential energy well and alters the energy deposition needed to escape the planet's gravity \citep{ErkaevEt2007A&A}. $R_H$ is the 
Hill sphere of the planet. 
For in-depth discussion of this formalism, refer to prior work \citep[e.g.,][]{Etangs2007A&A,ErkaevEt2007A&A,ValenciaEt2007bApJ,MurrayClayEt2009ApJ,Owen&Alvarez2015ApJ}. Note however, that the majority of the simulations performed here have $F_{\rm EUV} \la 10^4$ erg s$^{-1}$ cm$^{-2}$, in which case the mass is lost in the energy-limited regime where $P\rm{d}V$ work dominates the evaporation and the ionization fraction is low. We also restrict our study to planets with H/He-rich atmospheres where Equation 1 applies. We do not consider, for example, H/He diffusion escape processes that may be important for Earth-like atmospheric structures \citep[e.g.,][]{KastingEt2015ApJL}.

\begin{figure*}[t] 
\plotone{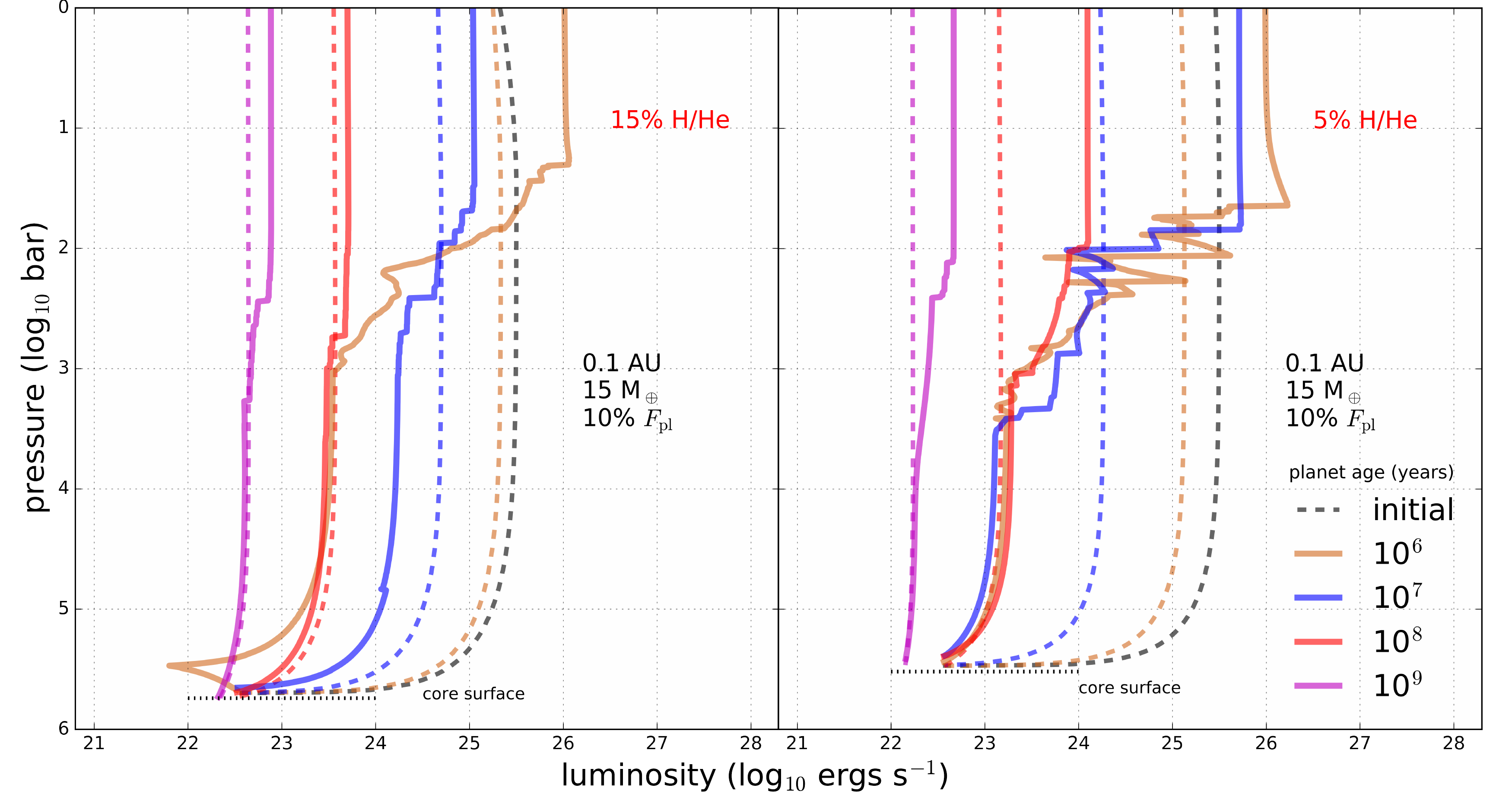}
\caption{\label{fig:profile}
Example pressure profiles of the H/He envelope for two 
planets at four different times showing the effects of planetesimal 
accretion at the level of $\fpl=10\%$ (solid lines). 
Both planets have initial $M_p/\mearth=15$, but the envelope masses are 
$F_g\equiv \menv/M_p=15$ (left) and $5\%$ (right). 
Equivalent thermal structures for the 
corresponding undisturbed planets (dashed lines) are also 
shown for comparison. The black dashed profile represents the initial 
thermal structure of the planets.
Clearly, planetesimal deposition perturbs a planet's internal thermal 
structure and can delay cooling by providing additional luminosity to 
the planet's interior. These effects are more significant for planets 
with a lower $F_g$, for a fixed $\fpl$ and $M_p$. 
As expected, the degree of perturbation decreases with time since last 
accretion. Yet, the accretion history do leave potentially 
measurable differences in the planetary structure compared to an 
undisturbed planet long ($\sim\,\gyr$) after the phase of high accretion rate.   
Additionally, there are small but continuous fluxes of minor impactors that maintains the resultant perturbations.}
\end{figure*}

\subsection{Planetesimal Accretion}
\label{method:pl}

The entry of large planetesimals into planetary interiors imparts substantial heating to the atmosphere. An example of this in the Solar System is the descent of the Shoemaker-Levy comet into Jupiter \citep{CrawfordEt1994,HammelEt1995}. While heating 
by the planetesimal is distributed over a large 
range in penetration depths, the maximum heating 
happens at the `airburst' location where the 
change of velocity of the planetesimal due to atmospheric drag is the maximum \citep{Zahnle+Mac1994Icarus}. This happens 
at a pressure where the impactor encounters a total atmospheric mass equal to its own mass. 
The pressure at airburst is given by

\begin{equation}
p_{\rm dep} \approx \sqrt{\frac{g^2 m_{\rm{pl}} \rho_{\rm{pl}}} {\pi C_d^2 H \sec(\phi)^3)}}
\end{equation}

where $g$ is the gravitational acceleration, $m_{\rm{pl}}$ and $\rho_{\rm{pl}}$ are the mass and mean density of the accreted planetesimal \citep[][]{Zahnle+Mac1994Icarus}. We assume all planetesimals to have a mean density comparable to those of Si-Fe (silicate-iron) mixtures, $\rho_{\rm{pl}}$ = 4 g/cm$^3$. $C_d$ is the drag coefficient; in this study, we assume all planestimals to have $C_d = 0.9$, accounting for their non-spherical geometries \citep[e.g.,][]{lagerros1998thermal}. $H\equiv kT/m_{\rm H}g$ is the atmospheric scale height that evolves with time and $\phi$ is the planetesimal's entry angle with respect to the tangential plane to the planet's surface. Since, for our choice of planet size and 
planet-star separation, the physical size of the 
planets is small compared to the size of the Hill 
sphere (e.g., $R_p/R_H\sim 0.05$ for $M_p/\mearth=10$ at $a=0.1\,\au$ around a Solar-mass star), planetesimal entry at 
the surface could be approximated as free-fall, i.e., the entry is radial ($\phi=90\degree$) and 
the velocity of the planetesimal at the surface is 
equal to the escape velocity $V_{\rm esc}$. 
We also assume isotropic accretion. 

For simplicity, we assume that the entirety of each planetesimal's kinetic energy $E_{\rm{pl}}\equiv 1/2 m_{\rm{pl}} V_{\rm esc}^2$ is liberated at an altitude corresponding 
to the airburst pressure for that planetesimal calculated using the instantaneous envelope structure of the planet at the time of accretion. 
We assume efficient heat transport (a compulsion 
due to the assumed spherical symmetry in {\sc mesa}) to distribute this liberated energy into the planet's atmosphere. In practice, we add this energy uniformly to the planet's zone layer corresponding to the airburst pressure. 
Properly following the mass of the planetesimal 
and the change in metallicity of the atmosphere 
due to the deposition of planetesimals is a hard 
task and depends intricately on the relevant 
chemistry of the planetesimal material. This is 
beyond the scope of this study. Instead, we add 
the mass of each planetesimal to the planet's core 
assuming that the 
typically heavier elements from the planestimals 
would ultimately 
sink to the bottom of the planet's envelope. 
We also leave the 
metallicity of the atmosphere unchanged for simplicity. 

The size distribution of planetesimals is assumed to be $dn/d\Rpl = \Rpl^{-3}$, where, $\Rpl$ is the planetesimal size \citep[e.g.,][]{Morbidelli+etal+2009Icarus}. We adopt 5 and 150 km as the lower and upper limits for $\Rpl$. This formalism is consistent with the mass and size distributions studied by state-of-the-art planetesimal formation algorithms (see e.g., \citealt{SimonEt2016ApJ}). 
Note that the lower and upper bounds in planetesimal sizes
are somewhat ad-hoc and guided by the motivation that all planetesimals should be low-mass ($\lesssim 10^{-6}\,\mearth$) compared to the planets and that a typical accretion event should be of very low-energy relative 
to the binding energy of the atmosphere. We will later (Section\ \ref{sec:results}) see that deeply penetrating (high-mass) planetesimals affect the envelope structure more. Thus the low lower-bound and steep power-law for the distribution of planetesimal sizes adopted in this study are conservative. Nevertheless, changes in these adopted values are unlikely to qualitatively alter the conclusions drawn from this study. 

We assume 
that the planetesimal accretion rate is 
flat in equal logarithmic time intervals 
between $10^5$ to $10^{10}$ years. 
This is motivated by the fact that successive instability timescales typically increase logarithmically \citep[e.g.,][]{chambers1996stability,funk2010stability,Chatterjee+etal+2008ApJ}. Furthermore, this choice allows us to not truncate accretion at an 
ad-hoc time, but at the same time, allows very low levels of accretion at late times (e.g., $\lesssim1/\yr$ by $1\,\myr$ and $\lesssim0.1/\yr$ by $10\,\myr$ for $\fpl=5\%$). 
We also consider a variant of this fiducial case in a handful of models where we assume that the accretion rate is flat in logarithmic intervals but 
all accretion completes within $100\,\myr$ in order to investigate the sensitivity of our results on the accretion termination time.  

Our models are characterized by the total planet mass ($M_p$), H/He mass fraction ($F_g\equiv M_{\rm env}/M_p$), and the 
mass fraction for the total accreted planetesimals 
($\fpl\equiv \mpl/M_p$). We generate planetesimal 
sizes and accretion times based on the size-distribution and accretion rate mentioned above until $\fpl$ attains the desired 
value (we explore $\fpl = 0$--$10\%$). Unless otherwise stated, we set the maximum time-step in {\sc mesa} to be comparable to the thermal timescale of the planet model. 
The relevant initial properties of the modeled planets and the level of $\fpl$ are summarized in Table\ 1.  

\begin{figure}[h] 
\begin{center}
\plotone{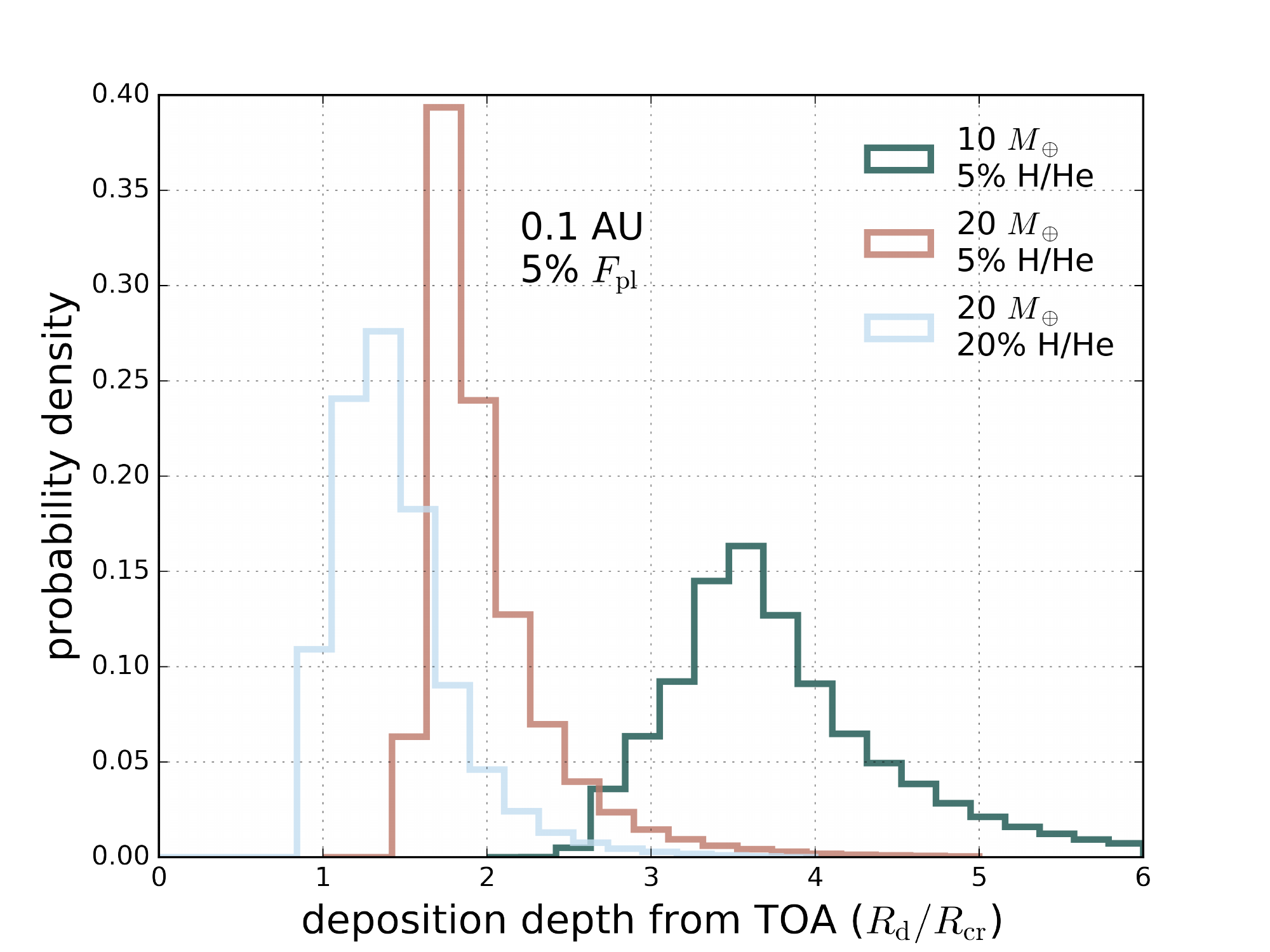}
\caption{\label{fig:rcr} Representative PDFs for the 
depths of planetesimal deposition with different initial
configurations of 
$M_p$ and $M_{\rm env}$. The horizontal axis shows the 
deposition depth from 
the top of the atmosphere (TOA) in units of the depth from 
the TOA for the radiative-convective boundary ($\rcr$). 
We find that planet models with lower $F_g$ or lower 
$M_p$ have deeper planetesimal depositions for any given 
$\fpl$. Although planets with lower $F_g$ (with fixed $M_p$) 
have larger radiative zones, the radiative-convective boundaries
are located at lower pressures. As a result, a higher fraction 
of planetesimals (for a given planetesimal size distribution) 
can penetrate deeper into the convective zone of the planets. 
}
\end{center}
\end{figure}  

\section{Results}
\label{sec:results}

Figure~\ref{fig:profile} shows example pressure profiles for 
two planets as a result of planetesimal accretion at different snapshots in time. Profiles of the same initial, but undisturbed planets are also shown for reference. Clearly, planetesimal accretion can alter the 
pressure profiles significantly. Depending on the time and 
the typical depth of planetesimal deposition, the 
changes may extend to large depths in the envelope. 
Furthermore, for the same initial planet mass and the total 
mass of accretion, 
the effects of planetesimal accretion are more pronounced in a 
planet with a lower gas-mass fraction 
($F_g$).
These deposited planetesimals effectively heat the deep 
interiors of the planet's envelope and lead to significantly 
delayed cooling. The delay in thermal cooling and additional 
luminosity due to accretion at deep interiors of the envelope 
can potentially lead to differences in the planet's observable 
properties even if the planets had similar initial properties. 

\begin{figure*}[t]
\begin{center}
\plotone{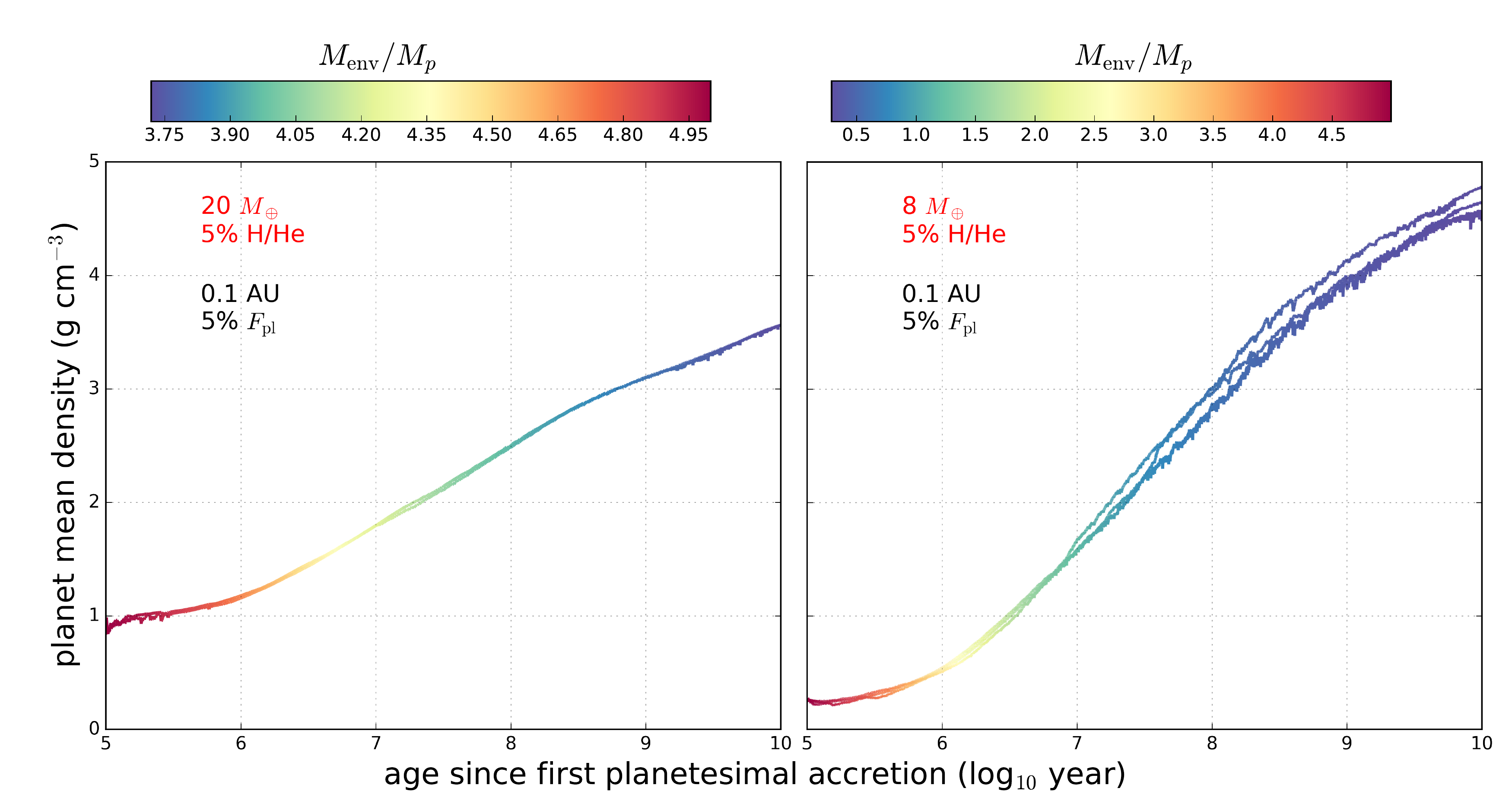}
\caption{\label{fig:density} Evolution of planetary 
mean density. Each panel shows the results of three 
simulations with the same initial planet properties 
and the same level of $\fpl$, differing only by the 
random draws of the planetesimal properties and their 
accretion times according to our adopted distributions 
(\S\ref{method:pl}). The colors indicate $F_g$ at any 
given time. 
The variations in the evolution trajectories and the final
densities come solely from the inherent stochastic 
variations in the planetesimal properties.  
This inherent stochasticity can lead to variations in 
mass-loss rates, especially at early times, and lead to 
variations in the final average density. The left and 
right panels show models for initial $M_p/\mearth = 20$ 
and $M_p/\mearth = 8$, respectively. All models have the 
same initial $F_g=5\%$ and are subjected to $\fpl=5\%$. 
The higher-mass planet shows a much lower level 
of stochastic variations for $t\gtrsim100\,\myr$. 
}
\end{center}
\end{figure*}

At a fundamental level, it is important where 
in the planet's envelope the planetesimals 
typically deposit their energy. 
Figure\ \ref{fig:rcr} shows the 
distributions of the deposition depth of 
accreted planetesimals normalized 
by the depth of the radiative-convective 
boundary ($\rcr$, from the top of the envelope) 
for three representative 
planets. 
We find that the airburst depth for a 
large fraction of planetesimals remain in the 
radiative zone of the planet's envelope (e.g.,
$M_p/\mearth=20$ with $F_g=20\%$ in 
Figure\ \ref{fig:rcr}). As $F_g$ decreases, more 
and more planetesimals are deposited in the 
convective zone of the envelope. Similarly, for 
the same $F_g$, lower-mass planets allow higher 
fraction of planetesimals to penetrate past the 
$\rcr$. 
As a consequence, the inflation of the planetary 
envelope due to the additional luminosity introduced via planetesimal accretion is less dramatic in higher-mass and higher-$F_g$ 
planets. As a result, the effects of 
planetesimal accretion by sub-Neptunes is 
expected to be more pronounced compared to those 
for the more well studied giant planets \citep{Leconte+Chabrier2012,Lozovsky2017jupiter}. 
Note that the exact distributions for the airburst depth for a given planet is subject to change if 
the distributions of planetesimal properties (Section\ \ref{method:pl}) are varied. However, the qualitative trends will remain the same.  

Higher-mass planetesimals deposit their energy
deeper into the 
planet's envelope. As a result, the effect of 
the deposition of a single high-mass 
planetesimal is different from the effect 
of the deposition of several low-mass 
planetesimals even when the total energy deposited by the low-mass planetesimals 
is equal to that deposited by the high-mass 
planetesimal. 
As a result, 
the effects of planetesimal accretion has a stochastic component 
strongly dependent on the typically low number of high-mass planetesimals 
(due to the mass function of planetesimals; 
Section\ \ref{method:pl}) and when they were accreted by the 
planet. Due to this inherent stochastic nature of the problem, 
the overall effect of planetesimal accretion can 
vary even when the initial $M_p$, $F_g$, and total mass fraction in accreted planetesimals 
($\fpl$) are all kept fixed simply due to 
statistical fluctuations. To quantify the 
degree of the 
stochasticity and how the stochastic 
fluctuations depend 
on the planet properties, we run two sets of simulations 
for two planets with 
slightly different initial composition and generate $3$ 
realizations each for the planetesimal masses and 
accretion times. 
The results are illustrated in Figure~\ref{fig:density}. 
We find that the stochastic effects are suppressed as the 
initial planet mass increases. For example, the evolution of 
the mean densities for the different realizations show little 
difference (except at very early times $t\la10^6$ year) 
for a planet with $M_p/\mearth=20$ (left panel, Figure\ \ref{fig:density}). Even with the same $F_g$ and $\fpl$, 
the stochastic effects are larger for a planet with $M_p/\mearth=8$ (right panel, Figure~\ref{fig:density}). 
In this lower-$M_p$ case, the three realizations of models 
show a variance in final density by ${\sim} 7\%$ starting 
from the same initial planet suffering the same level of 
planetesimal accretion simply due to statistical 
variations in the accretion history. 
As we will see, this result has important consequences when $\fpl$ also is varied.

\begin{figure*}[t]
\begin{center}
\plotone{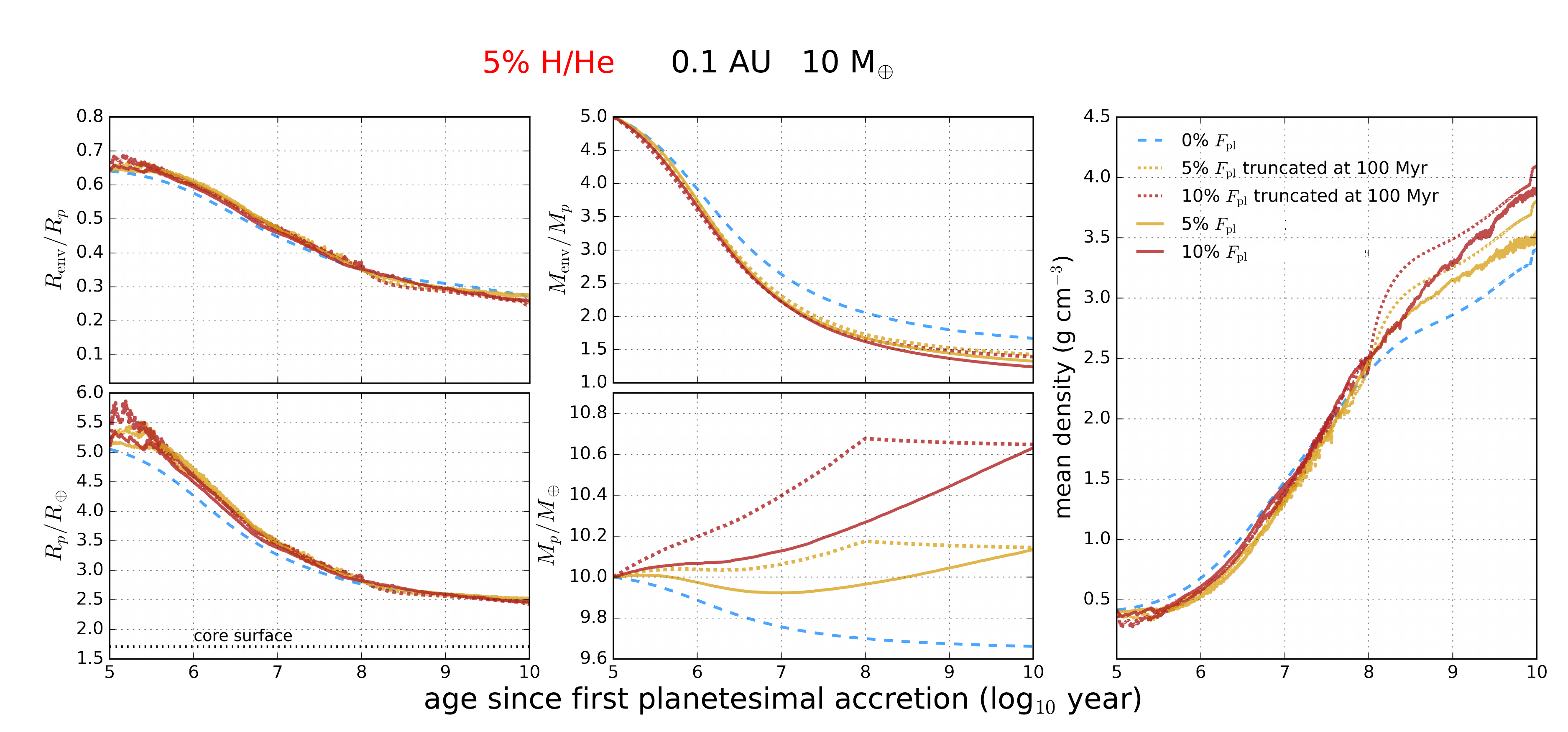}
\caption{\label{fig:evo10mp} Evolution of the structural 
properties as a result of planetesimal accretion with 
$\fpl=0$ (blue-dashed), $5$ (yellow), and $10\%$ (brown) for a planet with initial $M_p/\mearth=10$, 
$F_g=5\%$, at $a/\au=0.1$ around a Sun-like star. 
Dotted and solid lines denote cases where accretion is 
truncated at $100\,\myr$ since first accretion, and where 
no truncation is used (our fiducial case). 
Left, middle, and right panels show the evolution of the 
radius, mass, and density, respectively. The top and bottom 
panels show the properties of the envelope and the planet as 
a whole. The mass in the envelope always decreases 
due to photoevaporation, the amount of which is directly 
dependent on the level of $\fpl$. The mass of the planet can 
increase or decrease depending on whether the increase in core 
mass due to planetesimals is more than the loss of mass from 
the envelope or not.
During the early high-accretion phase, the planet's 
radius and the size of the envelope slightly
increase as a result of planetesimal accretion. The difference 
in the planet size between the disturbed and undisturbed cases 
reduce as the accretion rate drops over time 
(\S\ref{method:pl}). However, the enhanced evaporation from 
the envelope at early times leaves the eventual planet 
with a lower envelope mass than the undisturbed counterpart. 
The envelope loss combined with the increase in core mass 
over time, results in a lower final radius and higher mean 
density as $\fpl$ increases. Accretion-induced enhancement of 
envelope loss at early times is more severe if the same level of accretion happens 
within the first $100\,\myr$ in contrast if the accretion is spread out throughout the 
whole simulation time of $10\,\gyr$. As a result, the difference in final mean density 
relative to the undisturbed planet also is higher in the cases where accretion 
is truncated at $100\,\myr$.   
}
\end{center}
\end{figure*}

\begin{figure}[t]
\begin{center}
\plotone{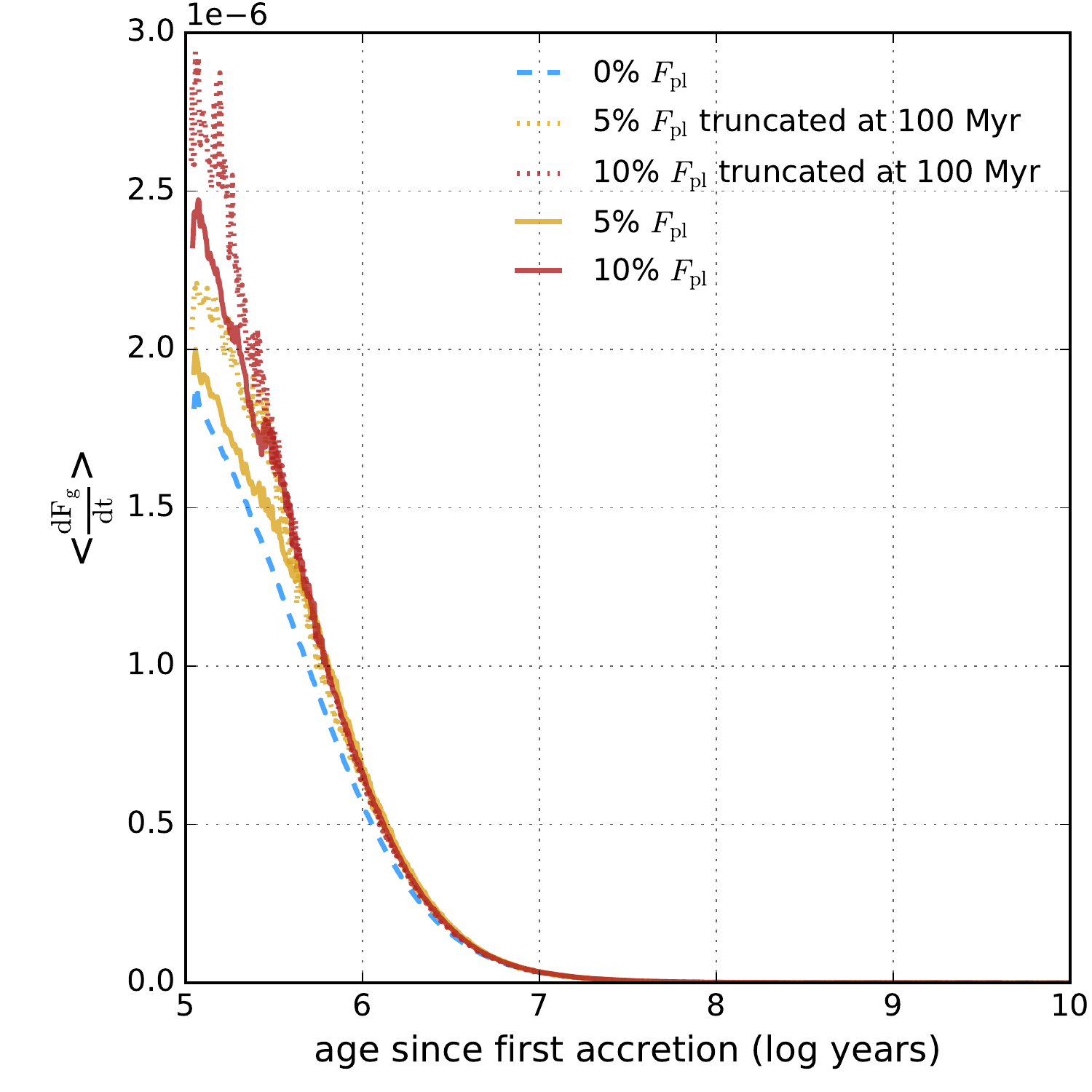}
\caption{\label{fig:dFgdt} Evolution of $dF_g/dt$ for a planet with initial 
$M_p/\mearth=10$, $F_g=0.05$, at $a/\au=0.1$ from a Solar-mass 
star as a result of various levels and history of planetesimal accretion. 
Line-styles and colors are the same as Figure\ \ref{fig:evo10mp}.  
Note that the accretion-induced enhancement in the mass-loss rate 
from the envelope is significant only up to $t\sim10\,\myr$. Subsequent 
low rate of accretion (Section\ \ref{method:pl}) does not affect envelope mass 
loss significantly relative to the undisturbed case.   
}
\end{center}
\end{figure}

\begin{figure*}[t]
\begin{center}
\plotone{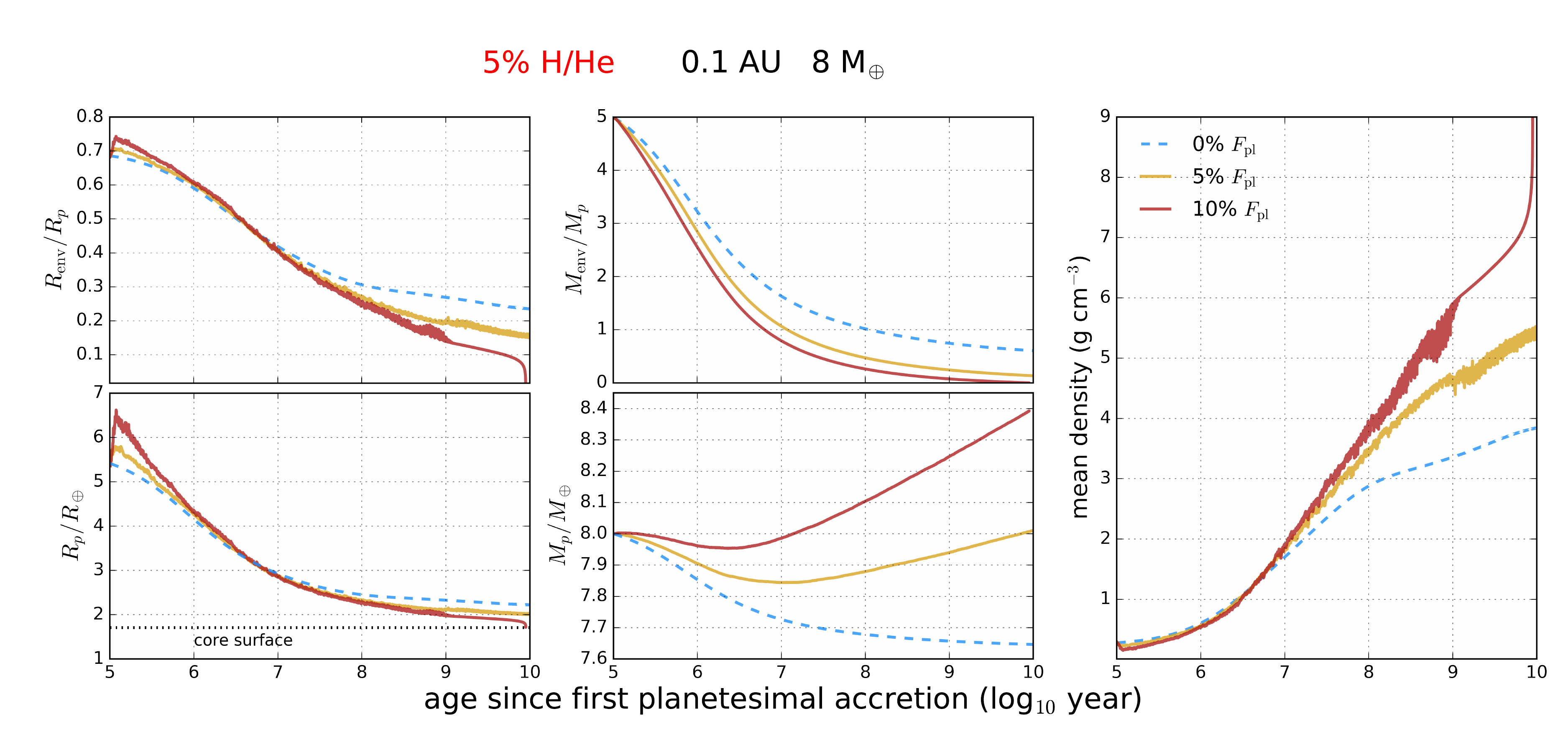}
\caption{\label{fig:evo8mp} Same as Figure\ \ref{fig:evo10mp} 
but for a planet with initial $M_p/\mearth=8$ and $F_g=5\%$. 
In this case, the 
enhanced mass loss from the envelope due to planetesimal accretion 
completely erodes the envelope for $\fpl=10\%$. Some envelope 
remains in the $\fpl=0$ and $5\%$ cases resulting in high variations 
in final average densities depending on $\fpl$. Note that 
the planet's evolution cannot be followed using {\sc MESA} when 
$F_g\lesssim 10^{-5}$. At this point we assume that whatever remains 
of the envelope quickly evaporates 
\citep[e.g., most recently][]{Owen+Wu2017arXiv} 
and the planet's density essentially is that of the core.  
}
\end{center}
\end{figure*}

\begin{figure*}[t]
\begin{center}
\plotone{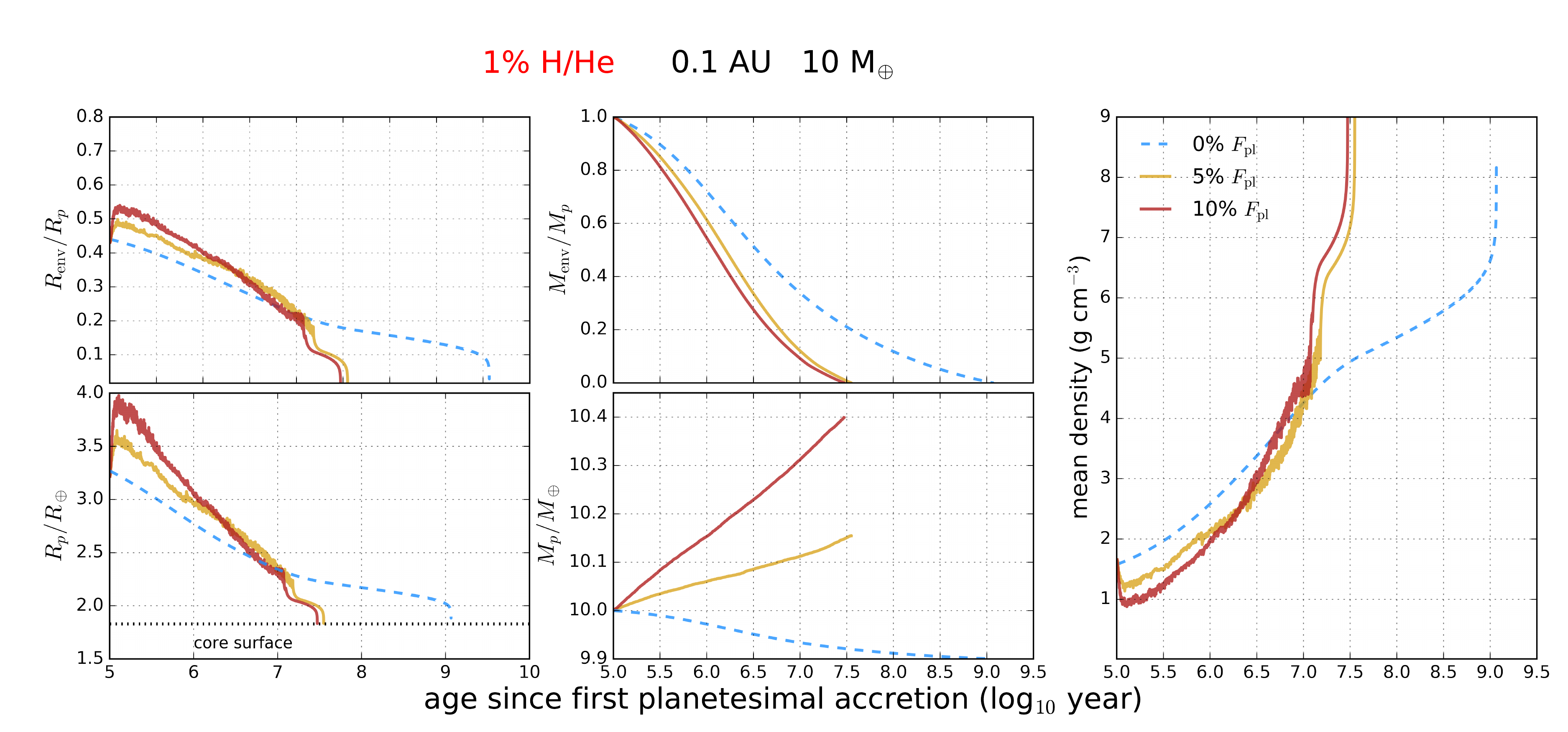}
\caption{\label{fig:evo10mp2} Same as Figure\ \ref{fig:evo10mp} 
but for a planet with initial $M_p/\mearth=10$ and $F_g=1\%$. 
In this case, the envelope of even the undisturbed planet is 
completely lost due to photo-evaporation. The level of $\fpl$ 
affects the planetary structures significantly while some fraction 
of the envelope is still left. Once the cores are naked, as 
expected, there is little difference in the observable properties 
of the planets subjected to different levels of planetesimal 
accretion. Note that once the envelope is fully stripped, we stop our integration and planetesimal accretion.  
}
\end{center}
\end{figure*}

We now focus on how variations in the level and history 
of planetesimal accretion can change the thermal evolution 
and final properties (including the mass and 
radius) of typical \kepler-like planets. 
Results from three sets of example simulations where 
we follow the effects of planetesimal accretion 
onto an evolving planet 
are illustrated in 
Figures\ \ref{fig:evo10mp}--\ref{fig:evo10mp2}. 
Each figure shows three simulations with the same initial 
planet properties, but modeled with varied 
levels ($\fpl=0, 5,$ and $10\%$) of 
planetesimal accretion. The most substantial 
difference due to planetesimal accretion 
in the evolution of the planet's 
structure comes due to the differences in the 
$F_g$ as a function of time. The additional 
luminosity due to planetesimal accretion 
expands the envelope and enhances mass loss 
from the envelope. The radius of the planet is 
{\it not} influenced substantially, however, depending 
on $M_p$, $F_g$ and $\fpl$, the mass loss from 
the envelope and the mass gain due to planetesimal accretion compete. Essentially, 
the low-density material from the envelope is 
compensated by high-density planetesimals as 
more and more planetesimals are accreted. As a 
result, the average densities of the planets are altered significantly. For 
example, a planet with initial $M_p/\mearth=10$, 
$F_g=5\%$, that accrets at the levels 
$\fpl=5$ and $10\%$, show $4\%$ and $20\%$ 
fractional differences in the average density relative to 
the undisturbed case at the integration stopping 
time of $t=10\,\gyr$ (Figure\ \ref{fig:evo10mp}).

The long-term ($\gtrsim\gyr$) differences in the mean density is entirely due to the 
differences in the envelope loss at early times ($\lesssim10\,\myr$) while the planetesimal accretion 
rate is high enough, and {\em not} due to transitional effects of recent individual accretion 
events. This is clearly illustrated in Figure\ \ref{fig:evo10mp}. 
To compare with our fiducial case of no termination of planetesimal accretion, 
we have simulated the same planet with the same levels of $\fpl$ and the same 
properties of planetesimals, but completed the accretion within $100\,\myr$. 
For any given $\fpl$, the difference between the final mean densities of the perturbed and undisturbed 
planets is higher in models where accretion is terminated at $t=100\,\myr$ compared to the difference  
exhibited in our fiducial models.  

The timescale to radiate away the additional energy from a particular accretion event is short. Thus, effects 
of particular accretion events are erased on short timescales. What really matters is the enhancement in a planet's size, 
and as a result, enhancement in the mass loss rate from the envelope due to 
planetesimal accretion during the early phase of high accretion rate. 
This is further illustrated in Figure\ \ref{fig:dFgdt} where we show the evolution of the average $dF_g/dt$ for various different 
levels of $\fpl$ with and without an imposed accretion termination time. In all cases, the mass loss rate 
from the envelope essentially converges with that for the undisturbed planet within $t\sim10\,\myr$. 
The enhanced envelope loss and core gain during this very early phase essentially set the final average properties of the planet. In the 
example case of truncated accretion, since the accretion rate during $t\lesssim10\,\myr$ is higher compared to the more 
conservative fiducial case, the enhancement in envelope loss is also higher, resulting in a higher difference 
in the mean density at late times. 

Since the difference in the average 
density of the planets at late times is 
primarily determined by the 
fraction of envelope retained by the planet at 
that time (for a given planet mass), 
the range in $M_p$ and $F_g$ where accretion can completely strip 
the envelope is of particular interest. 
In this regime, a sufficiently high level of $\fpl$ may completely strip 
the gaseous envelope\footnote{When the gas mass fraction is too low 
($F_g\lesssim 10^{-5}$), the 
atmosphere cannot be modeled using {\sc mesa} with mass-loss turned on. At this stage 
the code is halted. We assume that the remaining 
envelope is lost quickly leaving a naked core. }, 
whereas, lower levels of $\fpl$ would 
allow partial retention of the initial H/He envelope 
(Figure\ \ref{fig:evo8mp}). This can lead to 
even larger fractional difference in the average 
density of the planets due to varying levels 
$\fpl$ at the time of observation, even if the 
initial planet properties were the same. 
Of course, if the 
initial planet mass or the initial $F_g$ is 
sufficiently low, then photo-evaporation alone 
can strip the planet of its entire envelope. In this regime, 
independent of $\fpl$, the average 
density of the planet simply is that of the naked core. Thus in this 
regime, the observed densities would not be 
much different due to variations of $\fpl$ 
(Figure\ \ref{fig:evo10mp2}). However, even in 
this regime the inclusion of extra heat from accretion in the
interior of the planet dramatically offsets the
cooling contraction at young ages resulting in 
a runaway envelope loss at a much earlier time 
compared to the undisturbed planet.

Several inter-dependent processes are at work 
in producing these described behaviors in our
planet simulations. For example, the upturn in
planet radius during the epoch of high accretion
rate generally leads to a stronger mass-loss 
relative to an undisturbed planet. 
On the other hand, the increase in 
$M_{\rm core}/M_p$ due to the addition of the 
accreted planetesimals decreases the
gravitational scale-height of the atmosphere. 
The combination of lower $F_g$ and higher 
$M_{\rm core}/M_p$ leads to a lower planetary 
radii and higher average densities. This 
interplay between the increase in core mass 
and the amount of deep heating by planetesimals 
and the resulting envelope loss determines how 
the eventual mass-radius relation is shifted 
relative to an undisturbed planet.

\begin{figure}[h]
\begin{center}
\plotone{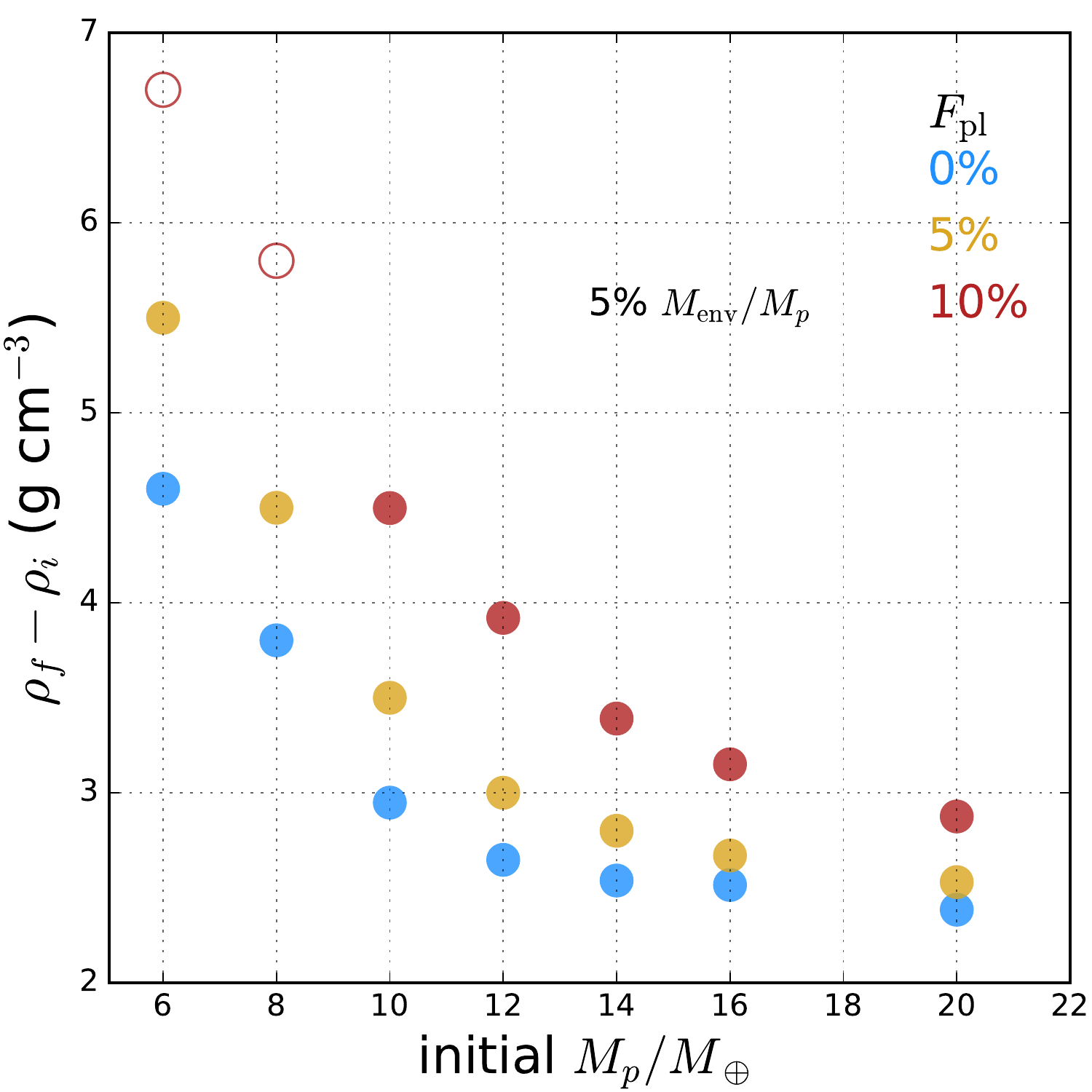}
\caption{\label{fig:rhovmp} Difference between the initial 
and final ($t=10\,\gyr$) average densities ($\Delta \rho_p$) 
as a function of the 
initial planet mass and a fixed initial $F_g=5\%$. For each 
$M_p$, three separate values of $\Delta \rho_p$ denotes 
$\fpl=0$ (blue), $5$ (yellow), and $10\%$ (brown). Open circles 
denote cases where the envelope completely evaporates before the 
integration stopping time of $10\,\gyr$. For the same level of 
initial $F_g$, $\Delta \rho_p$ decreases with increasing $M_p$ 
since lower mass planets have larger gravitational scale heights 
and suffers higher envelope loss for the same level of irradiation. 
In addition, the difference in magnitude of $\Delta \rho_p$ due 
to differences in $\fpl$ (as seen by the separation 
between the data-points for each $M_p$) also decreases. As $M_p$ 
increases, the fraction of planetesimals deposited deeper into the 
convective zone decreases (Figure\ \ref{fig:rcr}), reducing the
long-term effects of planetesimal accretion. }
\end{center}
\end{figure}

The time-dependent response, as well as the final structure of a planet due to 
planetesimal accretion depends strongly on the 
initial planet structure (for any given level 
of $\fpl$). We now focus on the structural 
differences of planets at $t\sim10\,\gyr$, which 
by design is long after the heavy accretion 
phase. Figure~\ref{fig:rhovmp} shows how 
a planet's average density at $t=10\,\gyr$
depends on the initial planet mass for the same 
initial $F_g$. In general, average density
increases over time simply due to
stellar-radiation driven mass-loss and thermal contraction. Planetesimal accretion enhances this 
mass loss and can lead to much higher final 
average densities at late times ($t=10\,\gyr$) 
depending on the initial planet properties. 
The overall mass-dependent trend for all levels 
of $\fpl$ is that the 
$\Delta \rho\equiv \rho_f-\rho_i$ decreases 
with increasing $M_p$. For the same initial $F_g$ 
a higher-mass planet has a smaller scale-height 
and for the same level of stellar irradiation 
suffers lower levels of mass loss from the 
envelope. Additional luminosity from planetesimal 
accretion enhances this mass loss. Moreover, 
the lower the $M_p$, the higher the effect of 
planetesimal accretion. Thus, the differences 
in $\Delta \rho$ between different levels of 
$\fpl$ increases with decreasing $M_p$. 

\begin{figure}[h]
\begin{center}
\plotone{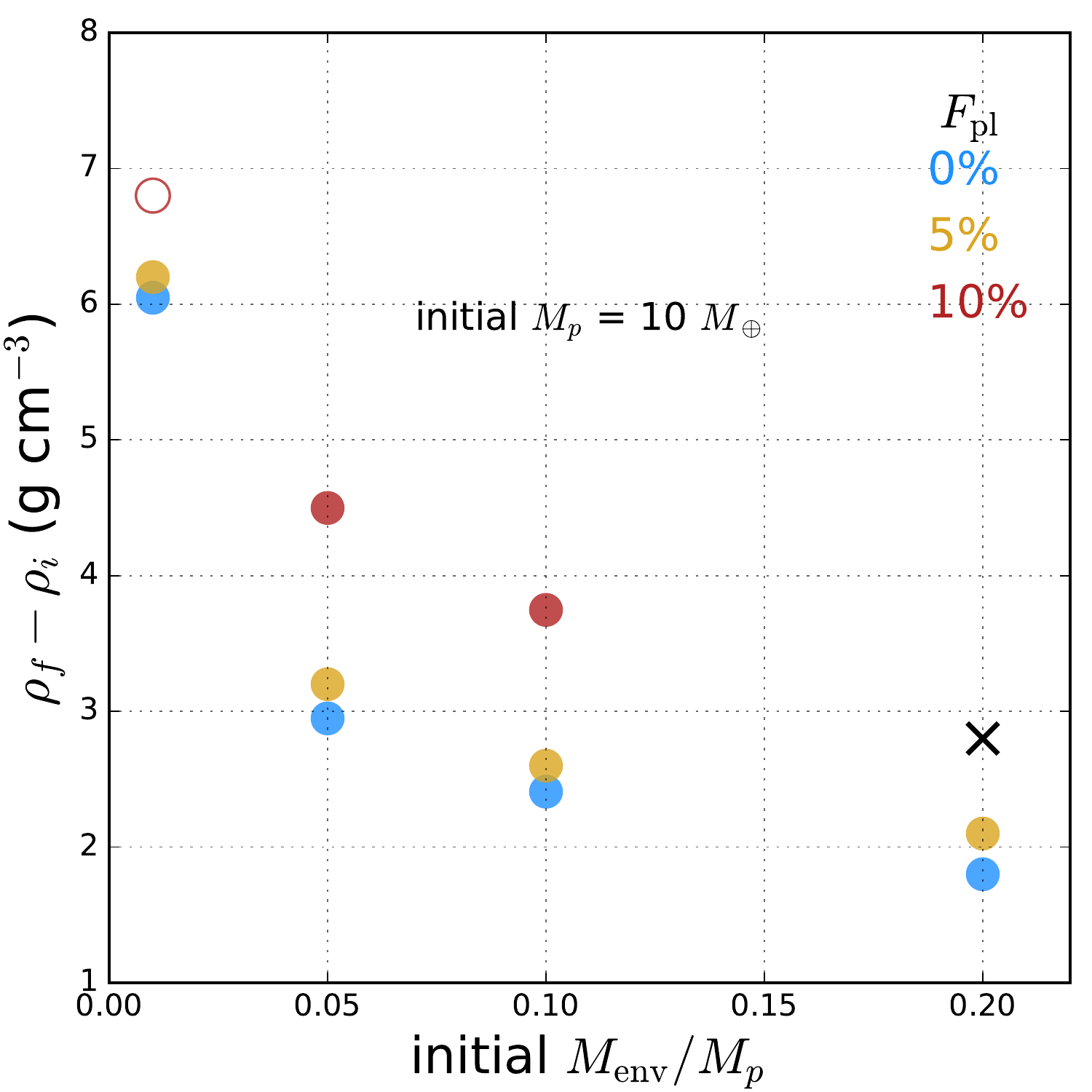}
\caption{\label{fig:rhovmenv} Same as Figure\ \ref{fig:rhovmp} 
but as a function of the initial $\menv/M_p$. The effects of 
planetesimal accretion is maximized for $\menv/M_p\sim5\%$ and 
falls off on either side. The cross denotes a failed model due to 
convergence issues. 
}
\end{center}
\end{figure}

For the same initial $M_p$, the effects of planetesimal 
accretion on $\Delta \rho$ due to varying levels of 
$\fpl$ has a more complicated dependence on the initial 
$F_g$ (Figure~\ref{fig:rhovmenv}). For very low initial $F_g$, even small levels of 
$\fpl$ can lead to a total loss of the envelope. Once 
the envelope is completely lost, of course the 
difference between the final densities and thus 
$\Delta \rho$
does not depend strongly on $\fpl$ anymore since 
at this point the average density is essentially 
the average density of the core. On the other hand, 
when the initial $F_g$ is sufficiently high, the effects 
of planetesimal accretion is low in general 
(Figure\ \ref{fig:rcr}), thus the differences 
in $\Delta \rho$ due to differences in $\fpl$ reduces. In our models we find that the maximum 
difference in $\Delta \rho$ is exhibited near 
$F_g\sim5$--$10\%$. 

\begin{figure*}[t]
\begin{center}
\plotone{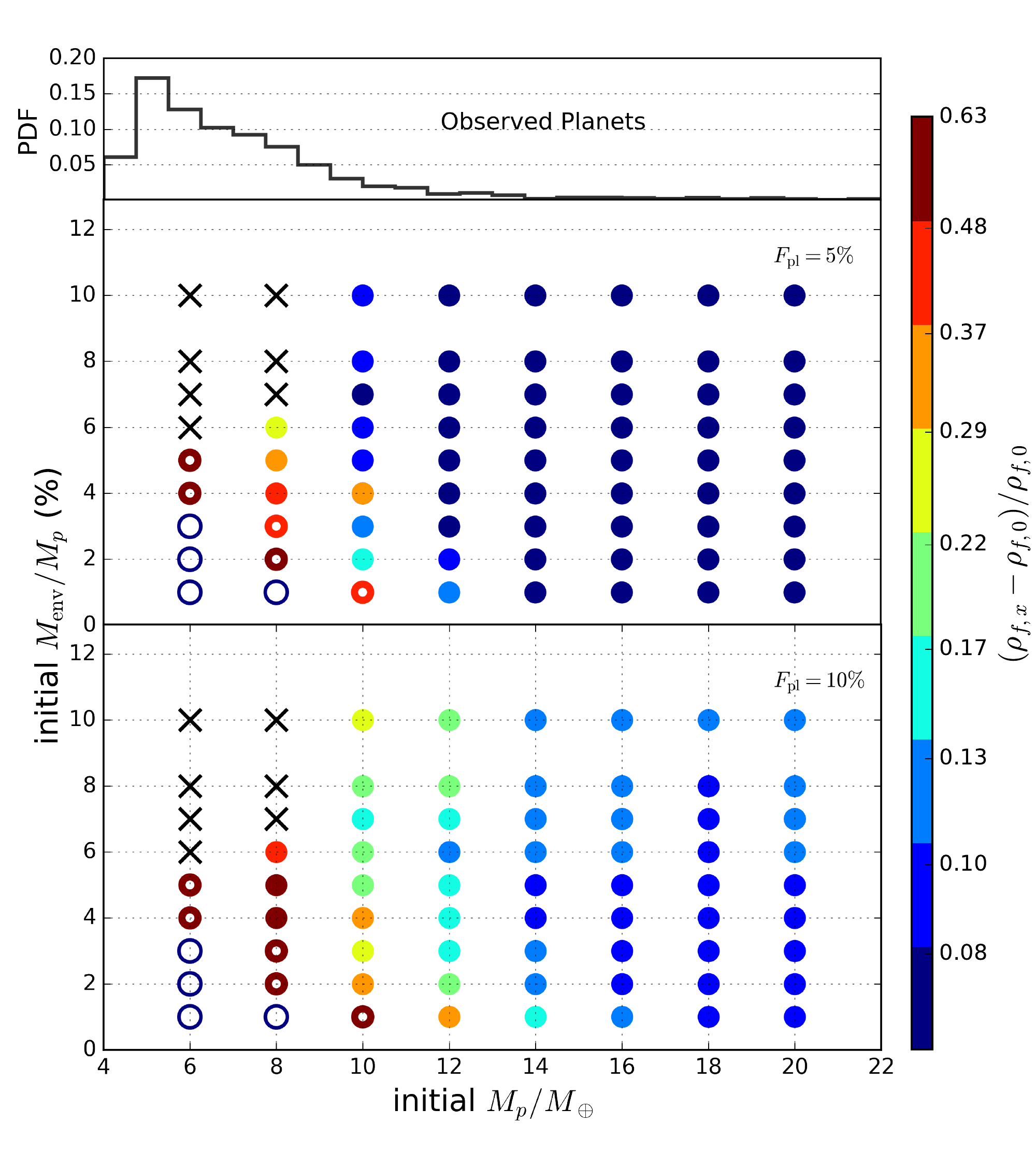}
\caption{\label{fig:rho_d} The fractional difference in the 
average density at 0.1 AU and $t=10\,\gyr$ (denoted by the colors) 
between undisturbed planets 
($\rho_{f,0}$) and those bombarded with planetesimals
($\rho_{f,x}$, where $x$ denotes $\fpl=x>0$), $\deltazerox$, as a function of the initial 
$M_p$ and $F_g$. Bottom and middle panels denote $\fpl=0.1$ 
and $0.05$, respectively. Filled circles denote 
models where some fraction of the envelope is retained in either 
$\fpl$ cases. Unfilled circles denote models where the envelope 
is completely eroded both in the $\fpl=0$ and $5\%$ (or $10\%$) cases. 
Partially filled circles denote models where the envelope is 
completely eroded in the $\fpl = 5$ or $\fpl=10\%$ case, but not for the
undisturbed planet. Crosses denote simulations that are 
numerically unstable (even without planetesimal accretion). 
For a fixed initial $F_g\gtrsim4\%$, as $M_p$ increases, 
$\deltazerox$ monotonically decreases with increasing initial 
$M_p$. The non-monotonic response, especially for low initial 
$M_p$ and $F_g$ can be explained by the fact 
that low mass planets (or planets with higher $F_g$) have more distended envelopes,
therefore are more susceptible to accretion-induced changes 
in the planetary radii. In addition, planets 
with lower $M_{\rm env}/M_p$ have deeper airburst depths for 
a given planetesimal mass (Figure\ \ref{fig:rcr}), thus the 
long-term structural properties are affected more. The top panel shows the distribution of observed planet mass for reference roughly estimated using $M_p/M_\oplus = 1.17 \times (R_p/R_\oplus)^{1.8}$ \citep{Chatterjee+Tan2015ApJL}. The regime where planetesimal accretion leads to the largest variety of final structures relative to undisturbed planets roughly coincides with the peak of the known exoplanets. }
\end{center}
\end{figure*}

This is likely better illustrated in 
Figure\ \ref{fig:rho_d} where we show 
the fractional difference in the average 
densities, $\deltazerox\equiv (\rho_{f,x}-\rho_{f,0})/\rho_{f,0}$,  
at $t=10\,\gyr$ 
between an undisturbed 
planet and a planet subjected to $\fpl=x$. Here, 
$\rho_{f,x}$ denotes the density at $t=10,\gyr$ for a planet 
subjected to $\fpl=x$. 
Figure\ \ref{fig:rho_d} shows results for $x=0.1$ and $0.05$ for a range in initial $M_p$ and $F_g$.  
For both levels of $\fpl$, for a given value of $F_g$, as initial 
$M_p$ increases, $\delta$ decreases. Similarly, for a 
given $M_p$, as $F_g$ decreases, $\deltazeroten$ and $\deltazerofive$ decrease, for $M_p/\mearth\gtrsim10$ and $M_p/\mearth\gtrsim16$, respectively. This trend reverses for lower-mass 
planets. Maximum $\deltazeroten$ is achieved 
for initially low-$M_p$ ($\lesssim12\,\mearth$) and moderately high-$F_g$ 
(between $0.01$ and $0.08$) planets. For $\deltazerofive$ the 
equivalent ranges are below initial $M_p/\mearth\lesssim16$ and 
any $0.01\lesssim F_g\lesssim0.08$. 
However, for 
$M_p/\mearth\lesssim10$, if initial $F_g$ is sufficiently low, 
then all of the envelope is lost even in an undisturbed case. 
Then $\delta$ for both $\fpl$ considered becomes very low. 
In the range of 
$M_p$ and $F_g$ we have considered in our models, and for the 
stellar irradiation on these models assuming a Sun-like star and 
a semi-major axis $a=0.1\,\au$, the highest $\delta$ we 
found at $t=10\,\gyr$ is $\approx60\%$. And this is achieved 
for planet models with initial $6\la M_p/\mearth \la 8$  
and $2\% \la F_g \la 5\%$. This regime of initially low-mass and
moderate-$F_g$ planets exhibit wide ranges in $\delta$ depending on the $\fpl$ and the details of the atmosphere retention fraction. 
Interestingly, in this regime the highest values of $\delta$ can coexist with the lowest values of $\delta$ depending on how easy it is to completely strip the initial envelope from a planet. The range of properties in this regime is very similar to those of the most common types of planets \kepler\ has discovered. Thus, depending on the accretion history, the most common types of exoplanets may achieve a wide range in structural properties at the time of observation even if they were created very similar.  

\section{Summary \& Discussion}
\label{sec:discussion}

This work represents the first study to incorporate the effects of
planetesimal accretion on thermally evolving planets with 
realistic models of planetary structures and including stellar 
irradiation. To keep the large parameter space manageable, we 
fix the planet-star distance at $a=0.1\,\au$, a typical distance 
for the planets found by \kepler. We also restrict our planet 
masses and mass fraction in the H/He envelope to values 
expected of sub-Neptunes, which represent the most common class 
of planets discovered (Section\ \ref{sec:meth}). We demonstrate that planetesimal accretion 
by sub-Neptunes post gas-disk dispersal can alter significantly
the evolutionary pathways taken by the planets primarily due 
to varying levels of envelope expansion due to the accretion-driven 
luminosity, and as a result, 
varying levels of envelope mass loss (Figures\ \ref{fig:evo10mp}--\ref{fig:evo10mp2}). 
We further show that if the initial $M_p$ and $F_g$ are within 
favorable ranges, differences in the accretion history and the 
resulting differences in the envelope mass loss can lead to 
differences in the observable properties including the average 
density of the planets long after the epoch of high rate of 
planetesimal accretion (Figures\ \ref{fig:rhovmp}--\ref{fig:rho_d}). 
These long-term differences are {\em not} due to recent events. Rather, 
these differences stem from the differences in enhanced expansion and 
envelope evaporation during the early epoch of high accretion rate. These early 
differences lead to differences in the final $F_g$ of the planet leading to observable 
differences in the mean density. As a result, if the same level of $\fpl$ is deposited with 
an early truncation in accretion (in contrast to our conservative fiducial case of no truncation; Section\ \ref{sec:meth}) 
the resulting higher initial accretion rate leads to even higher divergence in final $F_g$ and as a result 
final mean density of the planets relative to our fiducial case (Figure\ \ref{fig:evo10mp}).  

How the envelope responds to an accreted planetesimal depends on 
where in the envelope the planetesimal is deposited (and the amount of energy released at that location)- the deeper 
the deposition (and the more massive the planetesimal), the higher the impact on the structure of the 
envelope. In particular, the effects of planetesimal accretion 
is enhanced if a significant fraction of the 
planetesimals are deposited within the convective zone of the 
planet \citep[e.g.,][]{Lozovsky2017jupiter}. As a result, assuming that the size distribution of 
planetesimals is given by a power-law, the low number of deeply 
penetrating high-mass planetesimals can lead to stochastic 
differences in the planet's long-term observable properties, 
especially if the planet initially was low mass ($M_p/\mearth\lesssim10$) and had 
sufficient mass ($F_g\sim5$--$10\%$) in the envelope 
(Figure\ \ref{fig:density}). 
However, these 
stochastic effects are again suppressed if $F_g$ is too high 
since for a high-$F_g$ envelope, the radiative-convective 
boundary is at a higher pressures and for the same level and 
size-distribution of planetesimals, a lower fraction of 
planetesimals can penetrate deeper than the radiative-convective 
boundary (Figure\ \ref{fig:rcr}). Furthermore, as $F_g$ increases, 
the same level of response on the envelope at early times needs 
a higher accretion-driven luminosity. 

In general, the higher the mass fraction of planetesimal accretion, 
the higher the differences in the long-term observable properties 
for the planets. However, this trend is changed in the regime 
where the initial envelope is so tenuous that photo-evaporation over 
a few billion years alone 
can strip the planet of its envelope. In that regime, of course 
the final density of the planet is essentially dictated by the 
high-density core of the planet. Thus in this regime 
the level of planetesimal accretion does not lead to any significant 
difference in the average density of these planets (apart from the 
modest increase in the core mass due to planetesimals). Interestingly, the parameter space in $M_p$ and $F_g$ 
where the long-term observable 
properties differ by the maximum amount is actually adjacent to 
the above. Here high differences in, for example, the final 
average density is achieved because with high enough accretion 
the envelope is completely stripped whereas, a lower level of accretion may leave a significant fraction of the envelope 
keeping the planet puffy (Figure\ \ref{fig:rho_d}). 
This is particularly interesting since the most common types 
of exoplanets discovered by \kepler\ have properties very similar to this regime of high variability in final densities due to variations in the accretion history, even if they were formed with very similar properties (Figure\ \ref{fig:rho_d}). 

More specifically, we find that 
the evolution of planets born with $F_g\la 10\%$ and 
$4\lesssim M_p/\mearth\lesssim15$ are significantly altered due to
planetesimal accretion with $\fpl\gtrsim5\%$. 
Whereas, for $M_p/\mearth\lesssim10$ and $F_g\la 1\%$ the 
envelope is stripped entirely independent of the level of accretion 
leading to little differences in the final observable properties 
of these planets. In this same mass range, planets initially with 
$F_g\lesssim10\%$ are driven towards $F_g\sim0.5$--$1\%$ at 
$t\sim10\,\gyr$ (Table\ 1), in regime where the mass-loss timescales are 
maximized \citep{Chen+Rogers2016ApJ,Owen+Wu2017arXiv}. 
High-mass initial planets ($M_p/\mearth\gtrsim18$) 
typically are not significantly altered due to planetesimal accretion 
and follows very similar thermal evolution as their undisturbed 
counterparts driven primarily by the XUV irradiation. 

Since the effects of planetesimal accretion 
on a planet's thermal evolution depend 
strongly on the planet's initial structure, 
these effects are expected to be reflected 
in the observed planet population. Hence, 
planet formation or population synthesis studies, 
in addition to inputs from particular formation 
or evolution theories, 
should include accretion history as a crucial 
ingredient for modeling a planet's 
observable structure at late times, especially for 
present-day sub-Neptune-size planets. 
For instance, several studies propose that 
low-mass planets likely form with $F_g$ 
of only a few percent 
\citep{Ikoma&Hori2012ApJ,Lee&Chiang2016ApJ}.
\citet{Bodenheimer&Lissauer2014ApJ} also 
suggest a correlation between the core mass 
and the accreted mass of H/He. 
The thermal evolution of planets with low 
core as well as envelope mass is affected 
the most by any level of
planetesimal accretion. Furthermore, these 
are also the planets where the hysterisis 
from planetesimal accretion history is the 
most prominent. 
We strongly encourage follow-up studies 
to simulate a large population of planets 
with initial distribution of structural
properties guided by planet formation
theories \citep[e.g.,][]{Jin2017compositional} 
in order to investigate for long-term statistical 
trends due to planetesimal accretion on a 
population of young planets.

We point out that our models here
are restricted (to limit the enormous parameter 
space) to planets whose orbital
separations are kept fixed at $0.1\,\au$ 
in orbit to a Sun-like star, i.e., each 
planet model here was subjected to the 
same level of stellar irradiation. 
On one hand, a lower insolation flux would 
increase the fractional importance of the 
accretion-driven luminosity in the total 
energy budget and hence its 
effect. Thus it is expected that the 
thermal evolution of the planets subjected 
to planetesimal accretion would show 
a greater departure from the path taken by 
an undisturbed planet. Essentially, the 
envelope of a less irradiated planet can expand 
more before mass is stripped off it. Furthermore, 
a planet with a lower equilibrium temperature 
takes longer to radiate away the additional heat 
supplied by each accretion. This should lead 
to bigger variations in the observable
properties of young ($\sim1$--$10^2\,\myr$)
planets during the phase of high accretion
rate.  
On the other hand, a lower 
insolation flux also would necessitate 
a higher degree of envelope expansion for
any significant increase in the envelope
mass loss as a result of planetesimal
accretion. Since in the long term the observable 
differences are essentially created due to differences in 
the envelope mass loss (Figures\ \ref{fig:evo10mp}--\ref{fig:evo10mp2}), the memory of the history 
of planetesimal accretion would likely be less 
important after accretion rate decreases 
significantly. Thus a reduced variation in 
the observable properties of old 
($\sim10\,\gyr$) planets is expected. 

Somewhat related to the above is our 
assumption of no change in the accreting 
planet's semimajor axis as a result of 
planetesimal accretion. Earlier studies 
have shown that planetesimal scattering
could lead to changes in the planetary 
orbits, especially in near-resonant 
multiplanet systems 
\citep[e.g.,][]{murray1998migrating,Chatterjee+Ford2015ApJ}. The extent of planetesimal-driven migration should also be more important for relatively 
lower-mass planets 
\citep[most recently,][]{Chatterjee+Ford2015ApJ}. 
While, without the 
presence of a gas disk, planetesimal 
interactions at the level of a few to 
few ten percent is not expected to
result in large-scale orbital migration of the planets, planet-planetesimal
scattering typically 
reduces the eccentricity of the planet's 
orbit. As a result, the level of 
irradiation an initially eccentric planet 
is subjected to can change due to planet-planetesimal interactions 
simply due to the changes in the planet's 
orbital properties. 

Importantly, we note that the magnitudes of 
the structural changes of the planets driven by 
planetesimal accretion reported here are potentially 
the lower limits. Planetesimal deposition likely 
substantially enriches the envelopes with heavy 
elements and thereby enhances the metallicity. Indeed,
enhanced metallicity in the atmospheres of many
Neptune-sized planets have already been identified
\citep{bean2011optical,charnay20153d}. In this study 
we did not include the response of the envelope due 
to heavy-element enrichment (the Rosseland opacity 
tables used for this study do not include heavy 
metals). Higher metallicity may lead to 
non-negligible changes to planet radii simply by 
reducing the pressure scale height due to the 
increased mean molecular weight. However, the overall 
response is likely much more complicated. Increased 
metallicity can also increase the opacity of the 
envelope. Higher opacity would push the 
radiative-convective boundary upward making it easier 
for planetesimals to penetrate into the convective 
zone (see Figure~\ref{fig:rcr}) and affect the 
planet's envelope structure more efficiently. 
Furthermore, the energy released by a
planetesimal at some depth of the envelope would take 
longer to be radiated away, thus increasing the 
possibility of long-term hysterisis due to 
planetesimal accretion. We encourage more realistic 
models to study these interdependent effects. 
On a related note, the potential impact of late-stage 
accretion on atmospheric chemistry may also be 
important to consider and would depend on the 
exact chemical makeup of the deposited planetesimals.
For instance, \citet{MadhusudhanEt2016c} demonstrated
that ``pebble accretion" could directly enhance planet
metallicities and atmospheric C to O ratios shortly
after formation. Modeling chemical interactions 
between planetesimals and planetary atmospheres is
beyond the scope of our {\sc mesa} simulations, and
would require more complex 2D/3D atmospheric 
chemistry models and more rigorous mean opacity 
tables.

Finally, past studies have identified 
other energy sources that can alter a planet's 
envelope structure by supplying heat to the planet's 
interior. 
For example, the evolutionary changes in the host 
star may alter the FUV and X-ray flux and the 
resulting time-dependent heating of the planet's 
envelope \citep[e.g.,][]{cecchi2009relative}. 
Young and forming planets may be subjected to 
dramatic changes in the envelope structure due to 
giant impacts \citep[e.g.,][]{LiuEt2015ApJ}.  
If the forming planet's orbit has sufficient 
eccentricity, tidal heating may play a major role in 
determining the efficiency of gas accretion
\citep[e.g.,][]{Ginzburg+Sari2016}. \citet{Batygin&Stevenson2010ApJ} showed that if the 
planet's equlibrium temperature is sufficiently high 
($T_{\rm{eq}}\gtrsim1200\,\rm{K}$) heating via 
Ohmic dissipation in the envelope may be important. 
Many of the above processes require rather 
specific configurations or are effective at a specific 
stage of the evolution. In contrast, the planetesimal accretion mechanism may be ubiquitous. The 
core-accretion paradigm suggests that planets and 
planetesimals are both formed in the same disk and 
it is expected that planet-planetesimal interactions 
may be common post gas-disk dispersal 
\citep[see e.g.,][]{GoldreichEt2004}. Furthermore, the 
timescale for the onset of instability between the 
planets and planetesimals from a nearby disk may have 
wide ranges. Hence, different planets may have different 
time-since-last-accretion, enhancing the present-day variations 
in the observable properties. Thus, planet 
structure models, especially for the low-mass and 
low-$F_g$ planets typically found in the 
\kepler\ data, should include the effects of planetsimal accretion. 

\acknowledgements
We thank the anonymous referee for constructive comments. 
SC acknowledges support from CIERA through a
fellowship. HC acknowledges University Fellowship 
at Northwestern University; typically awarded to 
first year graduate students at NU. HC thanks Eve Lee and 
Pablo Marchant for helpful discussions. This research 
was supported in part through the computational resources and staff contributions provided for the Quest high performance computing facility at Northwestern University which is jointly supported by the Office of the Provost, the Office for Research, and Northwestern University Information Technology.

\software{MESA (v8845; Paxton et al. 2011, 2013, 2015)}

\clearpage

\startlongtable
\begin{deluxetable*}{cccc|c|cccc}
\label{T:props}
\tabletypesize{\scriptsize}
\tablecolumns{14}
\tablewidth{0pt}
\tablecaption{Summary of {\sc MESA} calculations.}
\tablehead{
    \multicolumn{4}{c}{Initial} &
    \colhead{} & 
    \multicolumn{4}{c}{Final (10~Gyr)} 
    \\
    \cline{1-4}
    \cline{6-9}
    \\
    \colhead{$M_p$} &
    \colhead{$R_p$} &
    \colhead{$F_g$} &
    \colhead{$\rho$} &
    \colhead{$\fpl$} &
    \colhead{$M_p$} & 
    \colhead{$R_p$} &
    \colhead{$F_g$} &
    \colhead{$\rho$} 
    \\
    \colhead{$(\mearth)$} &
    \colhead{$(\rearth)$} &
    \colhead{} & 
    \colhead{$(\rm{g\ cm}^{-3})$ } & 
    \colhead{} &
    \colhead{$(\mearth)$} &
    \colhead{$(\rearth)$} &
    \colhead{} & 
    \colhead{$(\rm{g\ cm}^{-3})$} 
    \\
}
\startdata
6 & 3.24 & 0.01 & 0.90 & 0 & -\tablenotemark{a} & - & - & - \\
6 & 3.73 & 0.01 & 0.86 & 5 & - & - & - & - \\
6 & 4.19 & 0.01 & 0.79 & 10 & - & - & - & - \\
6 & 4.02 & 0.02 & 0.47 & 0 & - & - & - & - \\
6 & 4.47 & 0.02 & 0.45 & 10 & - & - & - & - \\
6 & 4.70 & 0.03 & 0.30 & 0 & - & - & - & - \\
6 & 4.70 & 0.03 & 0.28 & 10 & - & - & - & - \\
8 & 3.79 & 0.02 & 0.78 & 0 & 7.81 & 2.05 & 0.001 & 5.12 \\
8 & 3.96 & 0.02 & 0.76 & 10 & - & - & - & - \\
8 & 4.38 & 0.03 & 0.6 & 0 & 7.79 & 2.08 & 0.003 & 4.87 \\
8 & 5.33 & 0.03 & 0.53 & 10 & - & - & - & - \\
8 & 4.83 & 0.04 & 0.38 & 0 & 7.78 & 2.10 & 0.003 & 4.62 \\
8 & 5.60 & 0.04 & 0.36 & 10 & - & - & - & - \\
8 & 5.30 & 0.05 & 0.29 & 0 & 7.65 & 2.25 & 0.007 & 3.69 \\
8 & 5.46 & 0.05 & 0.28 & 5 & 7.96 & 2.11 & 0.003 & 4.64 \\
8 & 6.33 & 0.05 & 0.27 & 10 & 8.28 & 1.95 & 0.001 & 6.16 \\
8 & 5.76 & 0.06 & 0.22 & 0 & 7.59 & 2.30 & 0.009 & 3.45 \\
8 & 5.97 & 0.06 & 0.22 & 10 & 8.21 & 2.09 & 0.002 & 4.99 \\
10 & 3.18 & 0.01 & 1.66 & 0 & - & - & - & - \\
10 & 3.33 & 0.01 & 1.62 & 5 & - & - & - & - \\
10 & 3.95 & 0.01 & 1.56 & 10 & - & - & - & - \\
10 & 3.73 & 0.02 & 1.04 & 0 & 9.84 & 2.25 & 0.005 & 4.74 \\
10 & 3.84 & 0.02 & 1.01 & 5 & 10.24 & 2.15 & 0.002 & 5.67 \\
10 & 4.12 & 0.02 & 0.97 & 10 & 10.65 & 2.09 & 0.001 & 6.41 \\
10 & 4.19 & 0.03 & 0.73 & 0 & 9.79 & 2.39 & 0.009 & 3.96 \\
10 & 4.41 & 0.03 & 0.71 & 10 & 10.59 & 2.25 & 0.005 & 5.10 \\
10 & 4.61 & 0.04 & 0.55 & 0 & 9.73 & 2.49 & 0.014 & 3.47 \\
10 & 5.01 & 0.05 & 0.43 & 0 & 9.67 & 2.58 & 0.018 & 3.11 \\
10 & 5.09 & 0.05 & 0.42 & 5 & 10.05 & 2.49 & 0.013 & 3.60 \\
10 & 5.20 & 0.05 & 0.42 & 10 & 10.48 & 2.48 & 0.013 & 3.77 \\
10 & 5.40 & 0.06 & 0.34 & 0 & 9.61 & 2.65 & 0.021 & 2.85 \\
10 & 5.58 & 0.06 & 0.33 & 10 & 10.41 & 2.56 & 0.016 & 3.41 \\
10 & 5.93 & 0.07 & 0.27 & 10 & 10.32 & 2.58 & 0.018 & 3.31 \\
10 & 6.15 & 0.08 & 0.23 & 0 & 9.46 & 2.76 & 0.028 & 2.48 \\
10 & 6.35 & 0.08 & 0.22 & 10 & 10.26 & 2.67 & 0.022 & 2.97 \\
10 & 6.89 & 0.1 & 0.16 & 0 & 9.30 & 2.84 & 0.033 & 2.23 \\
10 & 7.09 & 0.1 & 0.16 & 10 & 10.08 & 2.69 & 0.024 & 2.84 \\
12 & 3.22 & 0.01 & 1.92 & 0 & 11.90 & 2.21 & 0.002 & 6.05 \\
12 & 3.30 & 0.01 & 1.86 & 5 & 12.39 & 2.11 & 0.000 & 7.26 \\
12 & 3.47 & 0.01 & 1.84 & 10 & 12.89 & 2.08 & 0.000 & 7.92 \\
12 & 3.73 & 0.02 & 1.25 & 0 & 11.85 & 2.42 & 0.008 & 4.61 \\
12 & 3.77 & 0.02 & 1.23 & 5 & 12.34 & 2.37 & 0.006 & 5.12 \\
12 & 3.95 & 0.02 & 1.19 & 10 & 12.83 & 2.33 & 0.005 & 5.58 \\
12 & 4.15 & 0.03 & 0.91 & 0 & 11.81 & 2.57 & 0.014 & 3.86 \\
12 & 4.30 & 0.03 & 0.89 & 10 & 12.79 & 2.50 & 0.011 & 4.49 \\
12 & 4.53 & 0.04 & 0.70 & 0 & 11.75 & 2.68 & 0.020 & 3.35 \\
12 & 4.67 & 0.04 & 0.69 & 10 & 12.73 & 2.62 & 0.016 & 3.89 \\
12 & 4.89 & 0.05 & 0.56 & 0 & 11.70 & 2.79 & 0.025 & 2.98 \\
12 & 4.93 & 0.05 & 0.55 & 5 & 12.19 & 2.76 & 0.023 & 3.21 \\
12 & 5.00 & 0.05 & 0.54 & 10 & 12.68 & 2.73 & 0.021 & 3.43 \\
12 & 5.23 & 0.06 & 0.46 & 0 & 11.64 & 2.88 & 0.031 & 2.70 \\
12 & 5.35 & 0.06 & 0.45 & 10 & 12.63 & 2.85 & 0.028 & 3.02 \\
12 & 5.56 & 0.07 & 0.38 & 0 & 11.57 & 2.96 & 0.036 & 2.47 \\
12 & 5.68 & 0.07 & 0.37 & 10 & 12.56 & 2.90 & 0.031 & 2.85 \\
12 & 5.90 & 0.08 & 0.32 & 0 & 11.51 & 3.03 & 0.040 & 2.28 \\
12 & 6.53 & 0.1 & 0.23 & 0 & 11.36 & 3.16 & 0.049 & 1.99 \\
12 & 6.63 & 0.1 & 0.23 & 10 & 12.33 & 3.08 & 0.042 & 2.34 \\
14 & 3.23 & 0.01 & 2.22 & 0 & 13.91 & 2.34 & 0.003 & 6.00 \\
14 & 3.45 & 0.01 & 2.18 & 10 & 15.07 & 2.30 & 0.002 & 6.87 \\
14 & 3.72 & 0.02 & 1.47 & 0 & 13.86 & 2.54 & 0.010 & 4.65 \\
14 & 3.81 & 0.02 & 1.42 & 10 & 15.03 & 2.51 & 0.009 & 5.23 \\
14 & 4.12 & 0.03 & 1.08 & 0 & 13.82 & 2.69 & 0.017 & 3.90 \\
14 & 4.26 & 0.03 & 1.05 & 10 & 14.99 & 2.67 & 0.015 & 4.36 \\
14 & 4.48 & 0.04 & 0.84 & 0 & 13.77 & 2.82 & 0.024 & 3.38 \\
14 & 4.56 & 0.04 & 0.83 & 10 & 14.94 & 2.79 & 0.022 & 3.78 \\
14 & 4.82 & 0.05 & 0.68 & 0 & 13.72 & 2.94 & 0.031 & 2.99 \\
14 & 4.90 & 0.05 & 0.67 & 10 & 14.89 & 2.91 & 0.028 & 3.33 \\
14 & 5.14 & 0.06 & 0.56 & 0 & 13.67 & 3.04 & 0.037 & 2.69 \\
14 & 5.19 & 0.06 & 0.55 & 10 & 14.84 & 3.01 & 0.034 & 3.00 \\
14 & 5.44 & 0.07 & 0.47 & 0 & 13.61 & 3.13 & 0.044 & 2.44 \\
14 & 5.50 & 0.07 & 0.47 & 10 & 14.79 & 3.10 & 0.040 & 2.74 \\
14 & 5.74 & 0.08 & 0.40 & 0 & 13.55 & 3.22 & 0.050 & 2.24 \\
14 & 5.82 & 0.08 & 0.39 & 10 & 14.72 & 3.18 & 0.045 & 2.53 \\
14 & 6.30 & 0.1 & 0.30 & 0 & 13.42 & 3.38 & 0.061 & 1.92 \\
14 & 6.38 & 0.1 & 0.30 & 10 & 14.60 & 3.34 & 0.057 & 2.17 \\
16 & 3.25 & 0.01 & 2.50 & 0 & 15.91 & 2.43 & 0.005 & 6.09 \\
16 & 3.36 & 0.01 & 2.44 & 10 & 17.25 & 2.41 & 0.003 & 6.83 \\
16 & 3.72 & 0.02 & 1.66 & 0 & 15.87 & 2.64 & 0.012 & 4.78 \\
16 & 3.80 & 0.02 & 1.62 & 10 & 17.21 & 2.62 & 0.011 & 5.26 \\
16 & 4.10 & 0.03 & 1.25 & 0 & 15.83 & 2.79 & 0.020 & 4.00 \\
16 & 4.18 & 0.03 & 1.21 & 10 & 17.18 & 2.78 & 0.018 & 4.41 \\
16 & 4.46 & 0.04 & 0.98 & 0 & 15.79 & 2.93 & 0.027 & 3.47 \\
16 & 4.53 & 0.04 & 0.95 & 10 & 17.14 & 2.91 & 0.025 & 3.83 \\
16 & 4.77 & 0.05 & 0.80 & 0 & 15.75 & 3.05 & 0.035 & 3.06 \\
16 & 4.83 & 0.05 & 0.78 & 10 & 17.09 & 3.03 & 0.032 & 3.39 \\
16 & 5.07 & 0.06 & 0.67 & 0 & 15.70 & 3.16 & 0.042 & 2.74 \\
16 & 5.14 & 0.06 & 0.65 & 10 & 17.04 & 3.14 & 0.039 & 3.03 \\
16 & 5.40 & 0.07 & 0.56 & 10 & 17.00 & 3.23 & 0.045 & 2.77 \\
16 & 5.66 & 0.08 & 0.48 & 0 & 15.59 & 3.36 & 0.056 & 2.27 \\
16 & 5.76 & 0.08 & 0.47 & 10 & 16.94 & 3.33 & 0.052 & 2.53 \\
16 & 6.16 & 0.1 & 0.37 & 0 & 15.48 & 3.54 & 0.070 & 1.92 \\
16 & 6.20 & 0.1 & 0.37 & 10 & 16.84 & 3.50 & 0.065 & 2.17 \\
18 & 3.27 & 0.01 & 2.77 & 0 & 17.92 & 2.51 & 0.005 & 6.25 \\
18 & 3.35 & 0.01 & 2.71 & 10 & 19.42 & 2.50 & 0.005 & 6.87 \\
18 & 3.72 & 0.02 & 1.89 & 0 & 17.88 & 2.72 & 0.013 & 4.93 \\
18 & 3.80 & 0.02 & 1.86 & 10 & 19.39 & 2.71 & 0.012 & 5.36 \\
18 & 4.09 & 0.03 & 1.43 & 0 & 17.84 & 2.88 & 0.022 & 4.14 \\
18 & 4.24 & 0.03 & 1.39 & 10 & 19.35 & 2.87 & 0.020 & 4.51 \\
18 & 4.42 & 0.04 & 1.12 & 0 & 17.81 & 3.02 & 0.030 & 3.58 \\
18 & 4.50 & 0.04 & 1.10 & 10 & 19.32 & 3.01 & 0.027 & 3.92 \\
18 & 4.73 & 0.05 & 0.92 & 0 & 17.77 & 3.14 & 0.037 & 3.16 \\
18 & 4.79 & 0.05 & 0.90 & 10 & 19.28 & 3.13 & 0.035 & 3.46 \\
18 & 6.06 & 0.1 & 0.44 & 0 & 17.52 & 3.66 & 0.076 & 1.97 \\
18 & 6.10 & 0.1 & 0.44 & 10 & 19.06 & 3.62 & 0.071 & 2.21 \\
20 & 3.28 & 0.01 & 3.05 & 0 & 19.92 & 2.57 & 0.006 & 6.44 \\
20 & 3.32 & 0.01 & 3.00 & 5 & 20.76 & 2.57 & 0.006 & 6.76 \\
20 & 3.39 & 0.01 & 2.96 & 10 & 21.59 & 2.56 & 0.005 & 7.08 \\
20 & 3.72 & 0.02 & 2.11 & 0 & 19.89 & 2.78 & 0.014 & 5.10 \\
20 & 3.75 & 0.02 & 2.08 & 5 & 20.73 & 2.78 & 0.014 & 5.33 \\
20 & 3.83 & 0.02 & 2.04 & 10 & 21.56 & 2.77 & 0.013 & 5.61 \\
20 & 4.08 & 0.03 & 1.61 & 0 & 19.85 & 2.95 & 0.023 & 4.28 \\
20 & 4.19 & 0.03 & 1.57 & 10 & 21.54 & 2.93 & 0.021 & 4.70 \\
20 & 4.39 & 0.04 & 1.27 & 0 & 19.82 & 3.09 & 0.031 & 3.71 \\
20 & 4.48 & 0.04 & 1.25 & 10 & 21.50 & 3.08 & 0.029 & 4.07 \\
20 & 4.69 & 0.05 & 1.05 & 0 & 19.78 & 3.22 & 0.040 & 3.27 \\
20 & 4.72 & 0.05 & 1.04 & 5 & 20.63 & 3.21 & 0.038 & 3.44 \\
20 & 4.76 & 0.05 & 1.03 & 10 & 21.47 & 3.19 & 0.037 & 3.63 \\
20 & 4.96 & 0.06 & 0.89 & 0 & 19.74 & 3.34 & 0.048 & 2.93 \\
20 & 5.04 & 0.06 & 0.87 & 10 & 21.43 & 3.32 & 0.044 & 3.24 \\
20 & 5.30 & 0.07 & 0.75 & 10 & 21.40 & 3.42 & 0.052 & 2.94 \\
20 & 5.50 & 0.08 & 0.65 & 0 & 19.66 & 3.56 & 0.064 & 2.40 \\
20 & 5.55 & 0.08 & 0.65 & 10 & 21.35 & 3.53 & 0.060 & 2.68 \\
20 & 5.98 & 0.1 & 0.51 & 0 & 19.56 & 3.76 & 0.080 & 2.03 \\
20 & 6.03 & 0.1 & 0.50 & 10 & 21.26 & 3.72 & 0.074 & 2.29 \\
\enddata
\tablenotetext{a}{Omitted final properties denote numerically unstable models (even without planetesimal accretion). }
\end{deluxetable*}

\clearpage


\begin{thebibliography}{}
\expandafter\ifx\csname natexlab\endcsname\relax\def\natexlab#1{#1}\fi

\bibitem[{{Baraffe} {et~al.}(2003){Baraffe}, {Chabrier}, {Barman}, {Allard}, \&
  {Hauschildt}}]{BaraffeEt2003A&A}
{Baraffe}, I., {Chabrier}, G., {Barman}, T.~S., {Allard}, F., \& {Hauschildt},
  P.~H. 2003, \aap, 402, 701

\bibitem[{{Batygin} \& {Stevenson}(2010)}]{Batygin&Stevenson2010ApJ}
{Batygin}, K., \& {Stevenson}, D.~J. 2010, \apjl, 714, L238

\bibitem[{Bean {et~al.}(2011)Bean, D{\'e}sert, Kabath, Stalder, Seager,
  Kempton, Berta, Homeier, Walsh, \& Seifahrt}]{bean2011optical}
Bean, J.~L., D{\'e}sert, J.-M., Kabath, P., {et~al.} 2011, The Astrophysical
  Journal, 743, 92

\bibitem[{{Berardo} {et~al.}(2017){Berardo}, {Cumming}, \&
  {Marleau}}]{BerardoEt2017ApJ}
{Berardo}, D., {Cumming}, A., \& {Marleau}, G.-D. 2017, \apj, 834, 149

\bibitem[{{Bodenheimer} \& {Lissauer}(2014)}]{Bodenheimer&Lissauer2014ApJ}
{Bodenheimer}, P., \& {Lissauer}, J.~J. 2014, \apj, 791, 103

\bibitem[{{Bromley} \& {Kenyon}(2011)}]{Bromley+Kenyon+2011ApJ}
{Bromley}, B.~C., \& {Kenyon}, S.~J. 2011, \apj, 735, 29

\bibitem[{Cecchi-Pestellini {et~al.}(2009)Cecchi-Pestellini, Ciaravella,
  Micela, \& Penz}]{cecchi2009relative}
Cecchi-Pestellini, C., Ciaravella, A., Micela, G., \& Penz, T. 2009, Astronomy
  \& Astrophysics, 496, 863

\bibitem[{Chambers {et~al.}(1996)Chambers, Wetherill, \&
  Boss}]{chambers1996stability}
Chambers, J., Wetherill, G., \& Boss, A. 1996, Icarus, 119, 261

\bibitem[{Charnay {et~al.}(2015)Charnay, Meadows, Misra, Leconte, \&
  Arney}]{charnay20153d}
Charnay, B., Meadows, V., Misra, A., Leconte, J., \& Arney, G. 2015, The
  Astrophysical Journal Letters, 813, L1

\bibitem[{{Chatterjee} \& {Ford}(2015)}]{Chatterjee+Ford2015ApJ}
{Chatterjee}, S., \& {Ford}, E.~B. 2015, \apj, 803, 33

\bibitem[{{Chatterjee} {et~al.}(2008){Chatterjee}, {Ford}, {Matsumura}, \&
  {Rasio}}]{Chatterjee+etal+2008ApJ}
{Chatterjee}, S., {Ford}, E.~B., {Matsumura}, S., \& {Rasio}, F.~A. 2008, \apj,
  686, 580

\bibitem[{{Chatterjee} \& {Tan}(2015)}]{Chatterjee+Tan2015ApJL}
{Chatterjee}, S., \& {Tan}, J.~C. 2015, \apjl, 798, L32

\bibitem[{{Chen} \& {Rogers}(2016)}]{Chen+Rogers2016ApJ}
{Chen}, H., \& {Rogers}, L.~A. 2016, \apj, 831, 180

\bibitem[{Crawford {et~al.}(1994)Crawford, Boslough, Trucano, \&
  Robinson}]{CrawfordEt1994}
Crawford, D.~A., Boslough, M.~B., Trucano, T.~G., \& Robinson, A.~C. 1994,
  Shock waves, 4, 47

\bibitem[{{Erkaev} {et~al.}(2007){Erkaev}, {Kulikov}, {Lammer}, {Selsis},
  {Langmayr}, {Jaritz}, \& {Biernat}}]{ErkaevEt2007A&A}
{Erkaev}, N.~V., {Kulikov}, Y.~N., {Lammer}, H., {et~al.} 2007, \aap, 472, 329

\bibitem[{{Fang} \& {Margot}(2013)}]{Fang+Margot2013ApJ}
{Fang}, J., \& {Margot}, J.-L. 2013, \apj, 767, 115

\bibitem[{{Fernandez} \& {Ip}(1984)}]{Fernandez+Ip+1984Icarus}
{Fernandez}, J.~A., \& {Ip}, W.-H. 1984, \icarus, 58, 109

\bibitem[{{Freedman} {et~al.}(2014){Freedman}, {Lustig-Yaeger}, {Fortney},
  {Lupu}, {Marley}, \& {Lodders}}]{FreedmanEt2014}
{Freedman}, R.~S., {Lustig-Yaeger}, J., {Fortney}, J.~J., {et~al.} 2014, \apjs,
  214, 25

\bibitem[{{Freedman} {et~al.}(2008){Freedman}, {Marley}, \&
  {Lodders}}]{FreedmanEt2008}
{Freedman}, R.~S., {Marley}, M.~S., \& {Lodders}, K. 2008, \apjs, 174, 504

\bibitem[{Fulton {et~al.}(2017)Fulton, Petigura, Howard, Isaacson, Marcy,
  Cargile, Hebb, Weiss, Johnson, Morton, {et~al.}}]{FultonEt2017}
Fulton, B.~J., Petigura, E.~A., Howard, A.~W., {et~al.} 2017, \aj, 154, 109

\bibitem[{Funk {et~al.}(2010)Funk, Wuchterl, Schwarz, Pilat-Lohinger, \&
  Eggl}]{funk2010stability}
Funk, B., Wuchterl, G., Schwarz, R., Pilat-Lohinger, E., \& Eggl, S. 2010,
  Astronomy \& Astrophysics, 516, A82

\bibitem[{Ginzburg \& Sari(2016)}]{Ginzburg+Sari2016}
Ginzburg, S., \& Sari, R. 2016, Monthly Notices of the Royal Astronomical
  Society, stw2637

\bibitem[{{Goldreich} {et~al.}(2004){Goldreich}, {Lithwick}, \&
  {Sari}}]{GoldreichEt2004}
{Goldreich}, P., {Lithwick}, Y., \& {Sari}, R. 2004, \araa, 42, 549

\bibitem[{{Guillot} \& {Havel}(2011)}]{Guillot&Havel2011A&A}
{Guillot}, T., \& {Havel}, M. 2011, \aap, 527, A20

\bibitem[{{Hahn} \& {Malhotra}(1999)}]{Hahn+Malhotra+1999AJ}
{Hahn}, J.~M., \& {Malhotra}, R. 1999, \aj, 117, 3041

\bibitem[{Hammel {et~al.}(1995)Hammel, Beebe, Ingersoll, Orton,
  {et~al.}}]{HammelEt1995}
Hammel, H., Beebe, R., Ingersoll, A., Orton, G., {et~al.} 1995, Science, 267,
  1288

\bibitem[{Howe \& Burrows(2015)}]{Howe+Burrows2015ApJ}
Howe, A.~R., \& Burrows, A. 2015, The Astrophysical Journal, 808, 150

\bibitem[{{Howe} {et~al.}(2014){Howe}, {Burrows}, \& {Verne}}]{HoweEt2014ApJ}
{Howe}, A.~R., {Burrows}, A., \& {Verne}, W. 2014, \apj, 787, 173

\bibitem[{{Hwang} {et~al.}(2017){Hwang}, {Chatterjee}, {Lombardi}, {Steffen},
  \& {Rasio}}]{Hwang+Chatterjee+etal2017ApJ}
{Hwang}, J., {Chatterjee}, S., {Lombardi}, Jr., J., {Steffen}, J., \& {Rasio},
  F. 2017, arXiv:1707.01628

\bibitem[{{Ikoma} \& {Hori}(2012)}]{Ikoma&Hori2012ApJ}
{Ikoma}, M., \& {Hori}, Y. 2012, \apj, 753, 66

\bibitem[{{Inamdar} \& {Schlichting}(2015)}]{Inamdar+Schlichting2015}
{Inamdar}, N.~K., \& {Schlichting}, H.~E. 2015, \mnras, 448, 1751

\bibitem[{{Jackson} {et~al.}(2012){Jackson}, {Davis}, \&
  {Wheatley}}]{JacksonEt2012MNRAS}
{Jackson}, A.~P., {Davis}, T.~A., \& {Wheatley}, P.~J. 2012, \mnras, 422, 2024

\bibitem[{{JeongAhn} \& {Malhotra}(2017)}]{JeongAhn+Malhotra2017AJ}
{JeongAhn}, Y., \& {Malhotra}, R. 2017, \aj, 153, 235

\bibitem[{Jin \& Mordasini(2017)}]{Jin2017compositional}
Jin, S., \& Mordasini, C. 2017, arXiv preprint arXiv:1706.00251

\bibitem[{{Jin} {et~al.}(2014){Jin}, {Mordasini}, {Parmentier}, {van Boekel},
  {Henning}, \& {Ji}}]{JinEt2014ApJ}
{Jin}, S., {Mordasini}, C., {Parmentier}, V., {et~al.} 2014, \apj, 795, 65

\bibitem[{{Kasting} {et~al.}(2015){Kasting}, {Chen}, \&
  {Kopparapu}}]{KastingEt2015ApJL}
{Kasting}, J.~F., {Chen}, H., \& {Kopparapu}, R.~K. 2015, \apjl, 813, L3

\bibitem[{{Kasting} \& {Pollack}(1983)}]{Kasting+Pollack1983Icarus}
{Kasting}, J.~F., \& {Pollack}, J.~B. 1983, \icarus, 53, 479

\bibitem[{{Kirsh} {et~al.}(2009){Kirsh}, {Duncan}, {Brasser}, \&
  {Levison}}]{Krish+etal+2009Icarus}
{Kirsh}, D.~R., {Duncan}, M., {Brasser}, R., \& {Levison}, H.~F. 2009, \icarus,
  199, 197

\bibitem[{Lagerros(1998)}]{lagerros1998thermal}
Lagerros, J.~S. 1998, Astronomy and Astrophysics, 332, 1123

\bibitem[{{Lammer} {et~al.}(2003){Lammer}, {Selsis}, {Ribas}, {Guinan},
  {Bauer}, \& {Weiss}}]{LammerEt2003ApJ}
{Lammer}, H., {Selsis}, F., {Ribas}, I., {et~al.} 2003, \apjl, 598, L121

\bibitem[{{Lecavelier Des Etangs}(2007)}]{Etangs2007A&A}
{Lecavelier Des Etangs}, A. 2007, \aap, 461, 1185

\bibitem[{Leconte \& Chabrier(2012)}]{Leconte+Chabrier2012}
Leconte, J., \& Chabrier, G. 2012, Astronomy \& Astrophysics, 540, A20

\bibitem[{{Lee} \& {Chiang}(2016)}]{Lee&Chiang2016ApJ}
{Lee}, E.~J., \& {Chiang}, E. 2016, \apj, 817, 90

\bibitem[{Lehmer \& Catling(2017)}]{lehmer2017hydrodynamic}
Lehmer, O.~R., \& Catling, D.~C. 2017, The Astrophysical Journal, 845, 7pp

\bibitem[{{Liu} {et~al.}(2015){Liu}, {Hori}, {Lin}, \&
  {Asphaug}}]{LiuEt2015ApJ}
{Liu}, S.-F., {Hori}, Y., {Lin}, D.~N.~C., \& {Asphaug}, E. 2015, \apj, 812,
  164

\bibitem[{{Lopez} \& {Fortney}(2013)}]{Lopez&Fortney2013ApJ}
{Lopez}, E.~D., \& {Fortney}, J.~J. 2013, \apj, 776, 2

\bibitem[{{Lopez} {et~al.}(2012){Lopez}, {Fortney}, \&
  {Miller}}]{LopezEt2012ApJ}
{Lopez}, E.~D., {Fortney}, J.~J., \& {Miller}, N. 2012, \apj, 761, 59

\bibitem[{Lozovsky {et~al.}(2017)Lozovsky, Helled, Rosenberg, \&
  Bodenheimer}]{Lozovsky2017jupiter}
Lozovsky, M., Helled, R., Rosenberg, E.~D., \& Bodenheimer, P. 2017, The
  Astrophysical Journal, 836, 227

\bibitem[{Madhusudhan {et~al.}(2017)Madhusudhan, Bitsch, Johansen, \&
  Eriksson}]{MadhusudhanEt2016c}
Madhusudhan, N., Bitsch, B., Johansen, A., \& Eriksson, L. 2017, Monthly
  Notices of the Royal Astronomical Society, 469, 4102

\bibitem[{{Marcy} {et~al.}(2014){Marcy}, {Isaacson}, {Howard}, {Rowe},
  {Jenkins}, {Bryson}, {Latham}, {Howell}, {Gautier III}, {Batalha}, {Rogers},
  {Ciardi}, {Fischer}, {Gilliland}, \& {Kepler Team}}]{MarcyEt2014ApJS}
{Marcy}, G., {Isaacson}, H., {Howard}, A.~W., {et~al.} 2014, \apjs, 210, 20

\bibitem[{{Minton} \& {Levison}(2014)}]{Minton+Levison+2014Icarus}
{Minton}, D.~A., \& {Levison}, H.~F. 2014, \icarus, 232, 118

\bibitem[{{Moore} {et~al.}(2013){Moore}, {Hasan}, \&
  {Quillen}}]{Moore+etal+2013MNRAS}
{Moore}, A., {Hasan}, I., \& {Quillen}, A.~C. 2013, \mnras, 432, 1196

\bibitem[{{Morbidelli} {et~al.}(2009){Morbidelli}, {Bottke}, {Nesvorn{\'y}}, \&
  {Levison}}]{Morbidelli+etal+2009Icarus}
{Morbidelli}, A., {Bottke}, W.~F., {Nesvorn{\'y}}, D., \& {Levison}, H.~F.
  2009, \icarus, 204, 558

\bibitem[{{Mordasini} {et~al.}(2017){Mordasini}, {Marleau}, \&
  {Molli{\`e}re}}]{MordasiniEt2017arXiv}
{Mordasini}, C., {Marleau}, G.-D., \& {Molli{\`e}re}, P. 2017, ArXiv e-prints,
  arXiv:1708.00868

\bibitem[{{Mullally} {et~al.}(2015){Mullally}, {Coughlin}, {Thompson}, {Rowe},
  {Burke}, {Latham}, {Batalha}, {Bryson}, {Christiansen}, {Henze}, {Ofir},
  {Quarles}, {Shporer}, {Van Eylen}, {Van Laerhoven}, {Shah}, {Wolfgang},
  {Chaplin}, {Xie}, {Akeson}, {Argabright}, {Bachtell}, {Barclay}, {Borucki},
  {Caldwell}, {Campbell}, {Catanzarite}, {Cochran}, {Duren}, {Fleming},
  {Fraquelli}, {Girouard}, {Haas}, {He{\l}miniak}, {Howell}, {Huber}, {Larson},
  {Gautier}, {Jenkins}, {Li}, {Lissauer}, {McArthur}, {Miller}, {Morris},
  {Patil-Sabale}, {Plavchan}, {Putnam}, {Quintana}, {Ramirez}, {Silva Aguirre},
  {Seader}, {Smith}, {Steffen}, {Stewart}, {Stober}, {Still}, {Tenenbaum},
  {Troeltzsch}, {Twicken}, \& {Zamudio}}]{MullallyEt2015ApJS}
{Mullally}, F., {Coughlin}, J.~L., {Thompson}, S.~E., {et~al.} 2015, \apjs,
  217, 31

\bibitem[{Murray {et~al.}(1998)Murray, Hansen, Holman, \&
  Tremaine}]{murray1998migrating}
Murray, N., Hansen, B., Holman, M., \& Tremaine, S. 1998, Science, 279, 69

\bibitem[{{Murray-Clay} {et~al.}(2009){Murray-Clay}, {Chiang}, \&
  {Murray}}]{MurrayClayEt2009ApJ}
{Murray-Clay}, R.~A., {Chiang}, E.~I., \& {Murray}, N. 2009, \apj, 693, 23

\bibitem[{{Ormel} {et~al.}(2012){Ormel}, {Ida}, \&
  {Tanaka}}]{Ormel+etal+2012ApJ}
{Ormel}, C.~W., {Ida}, S., \& {Tanaka}, H. 2012, \apj, 758, 80

\bibitem[{{Owen} \& {Alvarez}(2015)}]{Owen&Alvarez2015ApJ}
{Owen}, J.~E., \& {Alvarez}, M.~A. 2015, \apj, 816, 34

\bibitem[{{Owen} \& {Wu}(2013)}]{Owen&Wu2013ApJ}
{Owen}, J.~E., \& {Wu}, Y. 2013, \apj, 775, 105

\bibitem[{{Owen} \& {Wu}(2017)}]{Owen+Wu2017arXiv}
---. 2017, \apj, 847, 29

\bibitem[{{Paxton} {et~al.}(2011){Paxton}, {Bildsten}, {Dotter}, {Herwig},
  {Lesaffre}, \& {Timmes}}]{PaxtonEt2011ApJS}
{Paxton}, B., {Bildsten}, L., {Dotter}, A., {et~al.} 2011, \apjs, 192, 3

\bibitem[{{Paxton} {et~al.}(2013){Paxton}, {Cantiello}, {Arras}, {Bildsten},
  {Brown}, {Dotter}, {Mankovich}, {Montgomery}, {Stello}, {Timmes}, \&
  {Townsend}}]{PaxtonEt2013ApJS}
{Paxton}, B., {Cantiello}, M., {Arras}, P., {et~al.} 2013, \apjs, 208, 4

\bibitem[{{Paxton} {et~al.}(2015){Paxton}, {Marchant}, {Schwab}, {Bauer},
  {Bildsten}, {Cantiello}, {Dessart}, {Farmer}, {Hu}, {Langer}, {Townsend},
  {Townsley}, \& {Timmes}}]{PaxtonEt2015ApJS}
{Paxton}, B., {Marchant}, P., {Schwab}, J., {et~al.} 2015, \apjs, 220, 15

\bibitem[{{Pu} \& {Wu}(2015)}]{Pu+Wu2015ApJ}
{Pu}, B., \& {Wu}, Y. 2015, \apj, 807, 44

\bibitem[{{Rogers}(2015)}]{Rogers2015ApJ}
{Rogers}, L.~A. 2015, \apj, 801, 41

\bibitem[{{Rogers} {et~al.}(2011){Rogers}, {Bodenheimer}, {Lissauer}, \&
  {Seager}}]{RogersEt2011ApJ}
{Rogers}, L.~A., {Bodenheimer}, P., {Lissauer}, J.~J., \& {Seager}, S. 2011,
  \apj, 738, 59

\bibitem[{{Rogers} \& {Seager}(2010)}]{Rogers&Seager2010bApJ}
{Rogers}, L.~A., \& {Seager}, S. 2010, \apj, 716, 1208

\bibitem[{{Rowe} {et~al.}(2014){Rowe}, {Bryson}, {Marcy}, {Lissauer},
  {Jontof-Hutter}, {Mullally}, {Gilliland}, {Issacson}, {Ford}, {Howell},
  {Borucki}, {Haas}, {Huber}, {Steffen}, {Thompson}, {Quintana}, {Barclay},
  {Still}, {Fortney}, {Gautier}, {Hunter}, {Caldwell}, {Ciardi}, {Devore},
  {Cochran}, {Jenkins}, {Agol}, {Carter}, \& {Geary}}]{RoweEt2014ApJ}
{Rowe}, J.~F., {Bryson}, S.~T., {Marcy}, G.~W., {et~al.} 2014, \apj, 784, 45

\bibitem[{Salz {et~al.}(2016)Salz, Schneider, Czesla, \&
  Schmitt}]{salz2016energy}
Salz, M., Schneider, P., Czesla, S., \& Schmitt, J. 2016, Astronomy \&
  Astrophysics, 585, L2

\bibitem[{{Saumon} {et~al.}(1995){Saumon}, {Chabrier}, \& {van
  Horn}}]{SaumonEt1995ApJS}
{Saumon}, D., {Chabrier}, G., \& {van Horn}, H.~M. 1995, \apjs, 99, 713

\bibitem[{Simon {et~al.}(2016)Simon, Armitage, Li, \& Youdin}]{SimonEt2016ApJ}
Simon, J.~B., Armitage, P.~J., Li, R., \& Youdin, A.~N. 2016, The Astrophysical
  Journal, 822, 55

\bibitem[{{Valencia} {et~al.}(2010){Valencia}, {Ikoma}, {Guillot}, \&
  {Nettelmann}}]{ValenciaEt2010A&A}
{Valencia}, D., {Ikoma}, M., {Guillot}, T., \& {Nettelmann}, N. 2010, \aap,
  516, A20

\bibitem[{{Valencia} {et~al.}(2007{\natexlab{a}}){Valencia}, {Sasselov}, \&
  {O'Connell}}]{ValenciaEt2007bApJ}
{Valencia}, D., {Sasselov}, D.~D., \& {O'Connell}, R.~J. 2007{\natexlab{a}},
  \apj, 665, 1413

\bibitem[{{Valencia} {et~al.}(2007{\natexlab{b}}){Valencia}, {Sasselov}, \&
  {O'Connell}}]{ValenciaEt2007ApJ}
---. 2007{\natexlab{b}}, \apj, 656, 545

\bibitem[{{Watson} {et~al.}(1981){Watson}, {Donahue}, \&
  {Walker}}]{WatsonEt1981}
{Watson}, A.~J., {Donahue}, T.~M., \& {Walker}, J.~C.~G. 1981, \icarus, 48, 150

\bibitem[{{Weiss} \& {Marcy}(2014)}]{Weiss&Marcy2014ApJL}
{Weiss}, L.~M., \& {Marcy}, G.~W. 2014, \apjl, 783, L6

\bibitem[{{Wolfgang} \& {Lopez}(2015)}]{Wolfgang+Lopez2015ApJ}
{Wolfgang}, A., \& {Lopez}, E. 2015, \apj, 806, 183

\bibitem[{{Wolfgang} {et~al.}(2016){Wolfgang}, {Rogers}, \&
  {Ford}}]{Wolfgang+Rogers+Ford2016ApJ}
{Wolfgang}, A., {Rogers}, L.~A., \& {Ford}, E.~B. 2016, \apj, 825, 19

\bibitem[{{Zahnle} \& {Mac Low}(1994)}]{Zahnle+Mac1994Icarus}
{Zahnle}, K., \& {Mac Low}, M.-M. 1994, \icarus, 108, 1

\end{thebibliography}

\end{document}